\documentclass[11pt]{article}

\usepackage{graphicx}% Include figure files
\usepackage{dcolumn}% Align table columns on decimal point
\usepackage{bm}% bold math
\usepackage{amssymb}
\usepackage{amsmath}
\usepackage{epsfig}    
\def\beq{\begin{equation}}
\def\eeq{\end{equation}}
\def\ba{\begin{array}}
\def\ea{\end{array}}
\def\bea{\begin{eqnarray}}
\def\eea{\end{eqnarray}}

\def\sq2{\sqrt{2}}
\def\End{\end{document}}
\newcommand{\smn}{\sigma^{\mu \nu}}

\newcommand{\sqd}{\sqrt{2}}

\allowdisplaybreaks

\title{Limits on the anomalous $Wtq$ couplings
}

\author{
R.\ Romero Aguilar, Antonio O.\ Bouzas\thanks{
abouzas@fis.mda.cinvestav.mx, 
corresponding author.}, F.\ Larios\\Departamento de F\'{\i}sica Aplicada,
CINVESTAV-M\'erida, A.P. 73, 97310 M\'erida,
Yucat\'an, M\'exico}

\begin{document}                                                              

\maketitle

\begin{abstract}
  Within the model-independent framework of $SU(3)\times SU(2)\times
  U(1)$ gauge-invariant dimension-six operators, we study
  flavor off-diagonal $Wtq$ couplings ($q=d,s$) and related four-quark
  contact interactions involving the top.  We obtain bounds on those
  couplings from Tevatron and LHC data for single-top production and
  branching fractions in top decays, as well as other experimental
  results on flavor-changing neutral-current processes including $B\to
  X_{q}\gamma$ and $Z\to b\bar{q}$ decays ($q=d,s$).  We also update
  the bounds on flavor-diagonal $Wtb$ couplings using the most recent
  measurement of $W$-helicity fractions in top decays from top-pair
  production.
%\pacs{\,14.65.Ha, 12.15.-y}
\end{abstract}

\maketitle

\section{Introduction}

Top-quark physics plays an essential role in the research program at
the LHC.  The top quark and the Higgs boson ---being the heaviest
known elementary particle and the only known elementary scalar,
respectively--- may be the best candidates to look for physics beyond
the Standard Model (SM) \cite{threviews}.  We may classify the
different studies on the top quark by the type of interactions they
consider, either flavor off-diagonal or diagonal. Within the class of
flavor-diagonal couplings we can find studies on $htt$
\cite{tophiggs}, $\gamma tt$ \cite{bouzas13,ttphoton}, $Ztt$
\cite{ttz}, $Gtt$ \cite{ttg}, $Wtb$
\cite{fabbrichesi,onofre,bach12,wtb}, as well as contact vertices such
as $ttqq$, $tbud$, $tb\nu e$ \cite{bach12,bach4,sarmiento,chen14}.
For the flavor off-diagonal case, we can find global studies that
include top couplings with several or all of the neutral gauge bosons
\cite{maltoni15,coimbra,topfcnc,zha14,zhu10,kamenik10}, as well as
more specific works on $htu(c)$ \cite{htu}, $\gamma tu(c)$
\cite{photontu}, $Ztu(c)$ \cite{ztu} and $Gtu(c)$ \cite{gtu}
couplings.  There are also studies on four-fermion interactions like
$tbff'$ and $td\nu e$ \cite{bach4,maltoni15}.  To date, there are no
similar studies on experimental limits for the flavor off-diagonal
charged-current (CC) $Wtq$ couplings available in the literature.

The main goal of this paper is to fill this gap by obtaining bounds on
flavor off-diagonal charged-current couplings of the top quark from
available experimental data.  We focus on the flavor off-diagonal
couplings $Wtd$ and $Wts$ as they arise in the basis of dimension-six
$SU(3)\times SU(2)\times U(1)$ gauge-invariant operators involving the
top quark.  We consider also contact four-quark interactions related
to the $Wtq$ couplings through the SM equations of motion.  In this
work we keep the flavor structure of the theory completely general, by
taking the dimension-six couplings as independent parameters.  Notice
that other theoretical flavor structures have been considered in the
literature, such as the Minimal Flavor Violation framework in which
the flavor mixing pattern of the SM is extended to the dimension-six
Lagrangian \cite{bach4,faller13}.  In addition to our analysis of the
flavor off-diagonal $Wtq$ vertices we also make an update on the
allowed parameter region of the flavor-diagonal $Wtb$ coupling, which
has received much attention in the recent literature
\cite{fabbrichesi,onofre,whelicity2,buckley}.  We asses the allowed
parameter regions for this vertex based on the cross sections for $tq$
and $tb$ production measured at the Tevatron and LHC, and the
measurement of $W$-helicity fractions in top decays from top-pair
production most recently reported \cite{cmswhelicity15}.

A minimal basis of $SU(3)\times SU(2)\times U(1)$ gauge-invariant
dimension-six operators involving the top quark has been given in
\cite{aguilaroperators,aguilar4}, and a complete one in \cite{rosiek}.
As far as top interactions are concerned those bases are identical,
aside from minor differences in the definition of contact four-fermion
operators.  We use that basis in this paper, as has become standard in
the recent literature. We carry out all computations at leading order
(LO) in both the SM and dimension-six effective couplings, fully
analytically in the case of decays and numerically with
\textsc{MadGraph5\_aMC@NLO} \cite{mg5} for scattering processes.
We adopt the operator normalization established in \cite{zha14} (and
references therein) at next-to-leading-order (NLO), which is
applicable also at LO and facilitates the counting of
coupling-constant powers, especially for automated computations. 

Due to $SU(2)\times U(1)$ gauge invariance and its spontaneous
breaking a complete separation of charged and neutral currents in
dimension-six operators is not possible.  As a consequence, most of
the basis operators involve interactions of both types in combinations
that may not be optimal to study a given process.  For those processes
we have to consider suitable linear combinations of basis operators
instead of the operators themselves. A similar strategy is used in
\cite{maltoni15}.  For those effective operators containing both CC
and neutral current (NC) terms, we take into account experimental data
for processes involving one or both types of vertices. Thus, besides
single-top production in hadron collisions (involving only flavor
off-diagonal charged-currents in the SM, but also flavor-changing
neutral currents (FCNC) in the
effective theory) and branching fractions in $t\to Wq$ decays, we
consider also FCNC vertices not involving the top such as $\gamma bq$
in $b\to d\gamma,s\gamma$ and $Zbq$ in $Z\to bd,bs$, as well as the
FCNC vertices in $t\to Zu,Zc$ and $pp(gu)\to t\gamma$ from
\cite{maltoni15}.  In this way, we survey the sensitivity of the
different processes to find the ones providing the best bounds for
each effective coupling.

The paper is organized as follows.  In section~\ref{sec:operators} we
list the dimension-six gauge-invariant operators relevant to this
study.  In section \ref{sec:fcnc} we analyze FCNC decay processes of
the top quark, the $Z$ boson and the $B$ meson, as well as flavor
off-diagonal top decays $t\to Wq$, that are used to obtain the limits
on the operators.  In section \ref{sec:single.top.prod} we discuss the
contribution of flavor off-diagonal effective operators to single-top
production at the Tevatron and the LHC.  In section \ref{sec:pheno} we
present the results obtained from the processes studied in the
previous sections on allowed regions for the $Wtd$ and $Wts$
effective couplings, the flavor-diagonal $Wtb$ couplings, and the
four-quark ones.  Finally, in section \ref{conclusions} we give our
conclusions.

\section{ Top quark dimension six operators}
\label{sec:operators}

New physics effects related to the top quark can be described
consistently by an effective electroweak Lagrangian that
satisfies the full $SU(3)_C\times SU(2)_L\times U(1)_Y$
gauge symmetry of the SM: 
\bea 
{\cal L} = {\cal
  L}_{\rm SM} + \frac{1}{\Lambda^2} \sum_k \left( C_k O^{(6)}_k + {\rm
    h.c.} \right) \; + \cdots\;,  \nonumber 
\label{su2lagrangian}
\eea 
where the ellipsis stands for operators of dimension higher than six.
$\Lambda$ is the scale of new, or beyond the SM physics.  The scale
$\Lambda$ is unknown but we will assume it to be $\Lambda = 1$ TeV as
is commonly used in the literature \cite{maltoni15,lambda}.  This is a
valid assumption given that the physical processes that are being
considered are at the significantly lower electroweak scale ($m_W$,
$m_t$ or $v$).  The Wilson coefficients $C_k$ depend on the scale, but
in tree level analyses this dependence is not taken into account
\cite{examples0}.  As experiments have reached higher precision it has
become appropiate to make studies at the next perturbative order,
where radiative corrections and renormalization dictate the dependence
of $C_k$ on the scale \cite{examples1}.  For instance, in
Ref.~\cite{zha14} we can find a study of top quark decay at NLO in QCD
where the operator mixing terms that appear at this level are taken
into account.  In particular, the W-helicity branching fractions of
$t\to bW$ decay at tree level only depend on $Wtb$ operators like
$O^{k3}_{uW}$ (defined below) but at NLO they can receive an indirect
contribution from the top-gluon operator $O_{qG}$ \cite{zha14}.
Nevertheless, our study is made at tree level for processes at (or
below) the top mass scale and we do not take into account the effects
of scale running and operator mixing.

Many years ago a long list of gauge invariant dimension-six operators
was introduced in Ref.~\cite{buchmuller86}.  Eventually, 
it was found that not all operators there are truly
independent \cite{wudka04,aguilaroperators}.  A revised list of
independent operators for the top-quark sector appeared first in
\cite{aguilaroperators,aguilar4}, and then a general revised list
for all the fields was provided in \cite{rosiek}.  Notice that
the list of top-gauge boson operators in \cite{aguilaroperators}
and in \cite{rosiek} coincide, except for the explicit notation
in a few cases (like $O^{ij}_{\phi \phi} \equiv O_{\varphi ud}^{ij}$).
From now on, we will refer
to the effective operators as defined in Ref.~\cite{rosiek}.
However, we adopt the sign convention in the covariant derivatives
as well as the operator normalization defined in \cite{zha14},
where a factor $y_t$ is attached to an operator for each
Higgs field it contains, and a factor $g$ ($g'$) for each $W$ ($B$)
field-strength tensor.    

As stated previously, we will follow the strategy of
Ref.~\cite{maltoni15}, where some of the operators considered there
are the same in our work. The original Lagrangian in
Eq.~(\ref{su2lagrangian}) is written in terms of gauge eigenstates but
we are referring to the physical (mass) eigenstates in our operators.
This means that additional Cabibbo-Kobayashi-Maskawa (CKM) suppressed
terms appear in the $Wtq$ vertices generated; for example, the
original diagonal operator $O_{\varphi q}^{(3)33}$ will generate a
$Wts$ non-diagonal coupling with a $V_{ts}$ factor.  We have taken
into account these mixing terms, but we point out that in the end
there is only a very small change in the allowed regions of
parameters.  Notice that there are recent studies on the potential of
the LHC to measure CKM matrix elements based on top quark rapidity
distribution \cite{aguilarckm}.  Our study is focused on the $Wtq$
vertices that originate in the dimension-six operators.

%%%%%%%%%%%%%%%%%%%%%%%%%%%%%%%%%%%%%%%%%%%%%%%%%%%%%%%%%%%%%%%%%%
\subsection{Effective $Wtq$ couplings of the top quark}
\label{wtqops}
%%%%%%%%%%%%%%%%%%%%%%%%%%%%%%%%%%%%%%%%%%%%%%%%%%%%%%%%%%%%%%%%%%%

Flavor indices aside, there are only four operators that give rise to
effective $Wtq$ couplings: $O_{\varphi q}^{(3)k3}$, $O_{\varphi
  ud}^{3k}$ (=$O^{3k}_{\phi \phi}$ in \cite{aguilaroperators}),
$O_{uW}^{k3}$ and $O_{dW}^{3k}$ ($k=1,2$).  The operator $O_{\varphi
  ud}^{3k}$ involves exclusively a charged-current vertex, but the
other three also generate NC couplings:
\begin{equation}
  \label{eq:orig}
\begin{aligned}
O_{\varphi q}^{(3)k3} &= 
\begin{aligned}[t]
&\frac{y_t^2}{2\sqrt{2}}g (v+h)^2
\left( W^+_\mu \bar u_{Lk} \gamma^\mu b'_L + W^-_\mu \bar d'_{Lk}
\gamma^\mu t_L \right)   \\
&+\frac{y_t^2}{2\sqrt{2}}\frac{g}{c_w} (v+h)^2 Z_\mu
(\bar u_{Lk} \gamma^\mu t_L - \bar d'_{Lk} \gamma^\mu b'_L)\,,  
\end{aligned} \\
O_{uW}^{k3} &= 
\begin{aligned}[t]
&2 y_t g (v+h) \left( \partial_\mu W^-_\nu 
+ ig W^3_\mu W^-_\nu \rule{0pt}{12pt}\right) \bar d'_{Lk} \smn t_R 
 \\
&+ \sqd y_t g (v+h) \left( c_W \partial_\mu Z_\nu +
s_W \partial_\mu A_\nu +
ig W^-_\mu W^+_\nu \right) \bar u_{Lk} \smn t_R \, ,    
\end{aligned}\\
O_{dW}^{3k} &= 
\begin{aligned}[t]
&2 y_t g (v+h) \left( \partial_\mu W^+_\nu 
+ ig W^+_\mu W^3_\nu \rule{0pt}{12pt}\right) \bar t_L \smn d_{Rk} 
 \\
&- \sqd y_t g (v+h) \left(c_W \partial_\mu Z_\nu +s_W \partial_\mu A_\nu +
ig W^-_\mu W^+_\nu \right) \bar b'_{L} \smn d_{Rk} \, .  
\end{aligned}
\end{aligned}
\end{equation}
With the aim of isolating NC of the up
quarks from those of the down quarks and of separating the $Z$ field
from the photon field $A$, we consider appropriate linear combinations
of $O_{\varphi q}^{(3)k3}$, $O_{uW}^{k3}$ and $O_{dW}^{3k}$ with the
purely NC operators $O_{\varphi q}^{(1)k3}$, $O_{uB}^{k3}$ and
$O_{dB}^{3k}$.  This strategy was also used in \cite{maltoni15}, where
the FCNC interactions of the top quark were analyzed.  We therefore
base our analysis on the following operators, written here in terms of
the physical vector boson fields:
\begin{align}
  \label{eq:operators}
&  \begin{aligned}
O_{\varphi q}^{(+)k3} &= O_{\varphi q}^{(1)k3}+O_{\varphi q}^{(3)k3}
=
\begin{aligned}[t]
&\frac{y_t^2}{2\sqrt{2}}g (v+h)^2
\left( W^+_\mu \bar u_{Lk} \gamma^\mu b'_L + W^-_\mu \bar d'_{Lk}
\gamma^\mu t_L \right)   \\
&- \frac{y_t^2}{2}\frac{g}{c_w} (v+h)^2 Z_\mu
\bar d'_{Lk} \gamma^\mu b'_L\,,  
\end{aligned}\\
O_{\varphi q}^{(-)k3} &= O_{\varphi q}^{(1)k3}-O_{\varphi q}^{(3)k3} 
= 
\begin{aligned}[t]
&-\frac{y_t^2}{2\sqrt{2}}g (v+h)^2
\left( W^+_\mu \bar u_{Lk} \gamma^\mu b'_L + W^-_\mu \bar d'_{Lk}
\gamma^\mu t_L \right)  \\
&- \frac{y_t^2}{4}\frac{g}{c_W} (v+h)^2 Z_\mu
\bar u_{Lk}\gamma^\mu t_{L} \, ,  
\end{aligned}\\ 
%  \end{aligned}\\
%&\addtocounter{equation}{-1} \begin{aligned}
O_{\varphi ud}^{3k} &= \frac{y_t^2}{2\sqd}g (v+h)^2
W^+_\mu \bar t_R \gamma^\mu d_{Rk} \,,
 \\
O_{uZ}^{k3} &= O_{uW}^{k3}-O_{uB}^{k3}
= 
\begin{aligned}[t]
&2 y_t g (v+h) \left( \partial_\mu W^-_\nu 
+ ig W^3_\mu W^-_\nu \rule{0pt}{12pt}\right) \bar d'_{Lk} \smn t_R 
 \\
&+ \sqd y_t g (v+h) \left( \frac{1}{c_W}\partial_\mu Z_\nu +
ig W^-_\mu W^+_\nu \right) \bar u_{Lk} \smn t_R \, ,    
\end{aligned}\\
O_{dZ}^{3k} &= O_{dW}^{3k}+O_{dB}^{3k}
= 
\begin{aligned}[t]
&2 y_t g (v+h) \left( \partial_\mu W^+_\nu 
+ ig W^+_\mu W^3_\nu \rule{0pt}{12pt}\right) \bar t_L \smn d_{Rk} 
 \\
&+ \sqd y_t g (v+h) \left( -\frac{1}{c_W}\partial_\mu Z_\nu +
ig W^+_\mu W^-_\nu \right) \bar b'_{L} \smn d_{Rk} \, ,  
\end{aligned}\\
O_{uA}^{k3} &= s^2_W O_{uW}^{k3}+ c^2_W O_{uB}^{k3}
= 
\begin{aligned}[t]
&2 y_t g s^2_W (v+h) \left( \partial_\mu W^-_\nu + ig
W^3_\mu W^-_\nu \rule{0pt}{12pt}\right) \bar d'_{Lk} \smn t_R 
 \\
&+ \sqd y_t g s^2_W (v+h) 
\left( \frac{1}{s_W}\partial_\mu A_\nu +
ig W^-_\mu W^+_\nu \right) \bar u_{Lk} \smn t_R \, ,    
\end{aligned} \\
O_{dA}^{3k} &= s^2_W O_{dW}^{3k}- c^2_W O_{dB}^{3k}
= 
\begin{aligned}[t]
&2 y_t g s^2_W (v+h) \left( \partial_\mu W^+_\nu 
+ ig W^+_\mu W^3_\nu \rule{0pt}{12pt}\right) \bar t_L \smn d_{Rk} 
 \\
&+ \sqd y_t g s^2_W (v+h) 
\left( -\frac{1}{s_W}\partial_\mu A_\nu +
ig W^+_\mu W^-_\nu \right) \bar b'_{L} \smn d_{Rk}\,.  
\end{aligned}
  \end{aligned}
\end{align}
Standard notation is used in this equation, with $I$, $J$, $K$ SU(2)
gauge indices, $\tau^I$ the Pauli matrices, and $\varphi$ the SM Higgs
doublet with $\tilde \varphi = i\tau^2 \varphi^*$.  The covariant
derivative is defined as $D_\mu \varphi = \partial_\mu \varphi - i g/2
\tau^I W^I_\mu \varphi - i g'/2 B_\mu \varphi$ \cite{zha14,maltoni15}.
The primed quark fields $d'$, $s'$, $b'$, are gauge eigenfields
related to mass eigenfields through the CKM matrix.  In equation
(\ref{eq:operators}), operators with $k=1,2$ yield flavor off-diagonal
effective $Wtq$ couplings, while those with $k=3$ correspond to
flavor-diagonal CC interactions.  (The latter have been considered in
\cite{fabbrichesi,onofre,whelicity2,buckley}.) From the point of view
of $Wtq$ interactions, the four operators $O_{\varphi q}^{(-)k3}$,
$O_{\varphi ud}^{3k}$, $O_{uZ}^{k3}$, $O_{dZ}^{3k}$ in
(\ref{eq:operators}) are completely equivalent to the original ones
$O_{\varphi q}^{(3)k3}$, $O_{\varphi ud}^{3k}$, $O_{uW}^{k3}$,
$O_{dW}^{3k}$ given in (\ref{eq:orig}) as listed in
\cite{aguilaroperators,rosiek}.  Notice that the operators
$O_{uA}^{k3}$ and $O_{dA}^{3k}$ contain the same CC vertices as
$O_{uZ}^{k3}$ and $O_{dZ}^{3k}$.  For the case of $O_{dA}^{3k}$, the
radiative decay $b\to q\gamma$ happens to be very sensitive to this
vertex as it contributes at tree level, and we will be able to show
limits of order $10^{-5}$ which are much stronger than any of the
other bounds.  We will then, neglect the potential effects of the CC
vertex of $O_{dA}^{3k}$ to single top production.  In the case of
$O_{uA}^{k3}$, as mentioned in \cite{maltoni15} the CMS measurement of
the parton level $gq\to t\gamma$ production process yields strong
constraints as well, and we will also be able to neglect its potential
effects.  This will also allow us to avoid $t$-channel photon-exchange
diagrams that would lead to divergent total cross sections.

The relations between the coefficients of the original operators
$O_{\varphi q}^{(1)k3}$, $O_{\varphi q}^{(3)k3}$, $O_{uW}^{k3}$,
$O_{dW}^{3k}$, $O_{uB}^{k3}$, $O_{dB}^{3k}$, and the new ones are
given by: \bea
\left(\begin{array}{c} C^{(+)k3}_{\varphi q} \\
    C^{(-)k3}_{\varphi q} \end{array}\right) &=& \frac{1}{2}
\left(\begin{array}{cc} 1 & 1
    \\
    1 & -1 \end{array}\right)
\left(\begin{array}{c} C^{(1)k3}_{\varphi q} \\
    C^{(3)k3}_{\varphi q} \end{array}\right),
\nonumber \\
\left(\begin{array}{c} C^{k3}_{uA} \\ C^{k3}_{uZ} \end{array}\right)
&=& \left(\begin{array}{cc} 1 & 1
    \\
    -s^2_W & c^2_W \end{array}\right) \left(\begin{array}{c}
    C^{k3}_{uB} \\ C^{k3}_{uW} \end{array}\right),
\nonumber \\
\left(\begin{array}{c} C^{k3}_{dA} \\ C^{k3}_{dZ} \end{array}\right)
&=& \left(\begin{array}{cc} -1 & 1
    \\
    s^2_W & c^2_W \end{array}\right) \left(\begin{array}{c}
    C^{k3}_{dB} \\ C^{k3}_{dW} \end{array}\right).  \nonumber \eea For
concreteness, in the rest of this paper we set $\Lambda \equiv 1$ TeV,
and write the dimensionful parameters in the operators in units of
TeV, namely, $v=0.246$, $m_t=0.1725$ and $m_W=0.0804$.  We will show
the limits on these coefficients below, but in addition we will
translate them to the limits on the form factors $V_{L(R)}$ and
$g_{L(R)}$ that are commonly used in the literature for the diagonal
$Wtb$ vertex \cite{fabbrichesi,onofre,aguilarwhel,cmswhelicitysingle}.
We will extend the definition to the flavor off-diagonal $Wtq$
couplings: $V^q_{L(R)}$ and $g^q_{L(R)}$.  The relation between the
form factors and the operator coefficients is given by:
\bea 
V^{q}_L &=& V_{tq}+ \frac{y^2_t}{2} \frac{v^2}{\Lambda^2}
\left( C_{\varphi q}^{(+)k3} - C_{\varphi q}^{(-)k3}\right)\, = V_{tq} +  
\left( C_{\varphi q}^{(+)k3} - C_{\varphi q}^{(-)k3}\right)/33.606 \, , 
\nonumber \\
V^{q}_R &=& \frac{y^2_t}{2} \frac{v^2}{\Lambda^2} C_{\varphi ud}^{3k} 
\, = C_{\varphi ud}^{3k}/33.606 \, , 
\nonumber \\
-g^{q}_R &=& \sqd gy_t\frac{v^2}{\Lambda^2} 
\left( C_{uZ}^{k3} +s_W^2 C_{uA}^{k3}\right) \,
= \left( C_{uZ}^{k3} +s_W^2 C_{uA}^{k3}\right)/18.156 \, ,
\nonumber \\
-g^{q}_L &=& \sqd gy_t \frac{v^2}{\Lambda^2}
\left( C_{dZ}^{3k} +s_W^2 C_{dA}^{3k}\right) \,
= \left( C_{dZ}^{3k} +s_W^2 C_{dA}^{3k}\right)/18.156 \, .
\label{factors}
\eea
Where $q=d(s)$ corresponds to $k=1(2)$.  The form factors
$V^{q}_{L(R)}$ and $g^q_{L(R)}$ for $q=b$ are equivalent to
$f_{1}^{L(R)}$ and $-f_{2}^{L(R)}$ in \cite{whelicity,whelicity2}.
Notice that the contributions by $C^{3k}_{uA}$ to
$g^q_{R}$ are suppresed by the $s^2_W$ factor, which is a desirable
feature as we are neglecting its effect on single-top production.

\subsection{Four-quark operators}
\label{sec:4q}

The choice of independent $Wtq$ operators in
\cite{aguilaroperators,rosiek} could have included another operator:
$O^{ij}_{qW}=\bar q_{Li} \gamma^\mu \tau^a D^\nu q_{Lj} W^a_{\mu\nu}$
that is associated with an off-shell $W$-propagator contribution to
single top quark production \cite{aguilar4,bach12}.  However, since
$O^{ij}_{qW}$ is related through the equations of motion to
four-fermion operators, we choose to include the latter in our basis
of independent operators.
Bases for the $SU(2)\times U(1)$-gauge invariant dimension-six
four-quark operators have been given in \cite{aguilar4,rosiek}.  The
four-quark operators related to $O^{ij}_{qW}$ involve left
quarks only.  In the notation of \cite{rosiek} the four-left-quark
basis operators are given by:
\begin{equation}
  \label{eq:4q.1}
  O_{qq}^{(1)ijkl} = (\overline{q}_{Li}\gamma_\mu q_{Lj})
                     (\overline{q}_{Lk}\gamma^\mu q_{Ll}),
\quad
  O_{qq}^{(3)ijkl} = (\overline{q}_{Li}\gamma_\mu \tau^{I}q_{Lj})
                     (\overline{q}_{Lk}\gamma^\mu \tau^{I}q_{Ll}).
\end{equation}
Other chiral structures can also contribute to single top
production, some of them have been considered in \cite{bach4}.
The operators (\ref{eq:4q.1}) involving the first and third families
that we consider in this paper are:
\begin{equation}
  \label{eq:4q.2}
  \begin{aligned}
    O_{qq}^{(1)1113} &= (\overline{u}_{L}\gamma_\mu
    u_{L}+\overline{d}'_{L}\gamma_\mu d'_{L}) (\overline{u}_{L}\gamma^\mu
    t_{L}+\overline{d}'_{L}\gamma^\mu b'_{L}), \\
    O_{qq}^{(3)1113} &= 2(\overline{u}_{L}\gamma_\mu
    d'_{L}) (\overline{d}'_{L}\gamma^\mu t_{L}) + 
    2(\overline{d}'_{L}\gamma_\mu
    u_{L}) (\overline{u}_{L}\gamma^\mu b'_{L}) + 
    (\overline{u}_{L}\gamma_\mu u_{L}-\overline{d}'_{L}\gamma_\mu
    d'_{L}) (\overline{u}_{L}\gamma^\mu
    t_{L}-\overline{d}'_{L}\gamma^\mu b'_{L}), \\
    O_{qq}^{(1)3113} &= (\overline{t}_{L}\gamma_\mu
    u_{L}+\overline{b}'_{L}\gamma_\mu d'_{L}) (\overline{u}_{L}\gamma^\mu
    t_{L}+\overline{d}'_{L}\gamma^\mu b'_{L}), \\
    O_{qq}^{(3)3113} &= 2(\overline{t}_{L}\gamma_\mu
    d'_{L}) (\overline{d}'_{L}\gamma^\mu t_{L}) + 
    2(\overline{b}'_{L}\gamma_\mu
    u_{L}) (\overline{u}_{L}\gamma^\mu b'_{L}) + 
    (\overline{t}_{L}\gamma_\mu u_{L}-\overline{b}'_{L}\gamma_\mu
    d'_{L}) (\overline{u}_{L}\gamma^\mu
    t_{L}-\overline{d}'_{L}\gamma^\mu b'_{L}), \\
    O_{qq}^{(1)1133} &= (\overline{u}_{L}\gamma_\mu
    u_{L}+\overline{d}'_{L}\gamma_\mu d'_{L}) (\overline{t}_{L}\gamma^\mu
    t_{L}+\overline{b}'_{L}\gamma^\mu b'_{L}), \\
    O_{qq}^{(3)1133} &= 2(\overline{d}'_{L}\gamma_\mu
    u_{L}) (\overline{t}_{L}\gamma^\mu b'_{L}) + 
    2(\overline{u}_{L}\gamma_\mu
    d'_{L}) (\overline{b}'_{L}\gamma^\mu t_{L}) + 
    (\overline{u}_{L}\gamma_\mu u_{L}-\overline{d}'_{L}\gamma_\mu
    d'_{L}) (\overline{t}_{L}\gamma^\mu
    t_{L}-\overline{b}'_{L}\gamma^\mu b'_{L}), \\
    O_{qq}^{(1)3313} &= (\overline{t}_{L}\gamma_\mu
    t_{L}+\overline{b}'_{L}\gamma_\mu b'_{L}) (\overline{u}_{L}\gamma^\mu
    t_{L}+\overline{d}'_{L}\gamma^\mu b'_{L}), \\
    O_{qq}^{(3)3313} &= 2(\overline{t}_{L}\gamma_\mu
    b'_{L}) (\overline{d}'_{L}\gamma^\mu t_{L}) + 
    2(\overline{b}'_{L}\gamma_\mu
    t_{L}) (\overline{u}_{L}\gamma^\mu b'_{L}) + 
    (\overline{t}_{L}\gamma_\mu t_{L}-\overline{b}'_{L}\gamma_\mu
    b'_{L}) (\overline{u}_{L}\gamma^\mu
    t_{L}-\overline{d}'_{L}\gamma^\mu b'_{L}),
  \end{aligned}
\end{equation}
with $O_{qq}^{(1,3)3113}$ and $O_{qq}^{(1,3)1133}$ Hermitian.  All of
the operators (\ref{eq:4q.2}) are relevant to single-top production,
except for $O_{qq}^{(1)3113}+O_{qq}^{(3)3113}$ and $O_{qq}^{(1)1133}$
which contain only terms with an even number of top fields and can be
bounded by their contribution to top-pair production.  The operators
\begin{equation}
\label{eq:sst}
  \begin{aligned}
O_{qq}^{(1)1313} &=
(\overline{u}_{L}\gamma_\mu t_{L}+\overline{d}'_{L}\gamma_\mu b'_{L})
(\overline{u}_{L}\gamma^\mu t_{L}+\overline{d}'_{L}\gamma^\mu
b'_{L}),\\
O_{qq}^{(3)1313} &=
4(\overline{u}_{L}\gamma_\mu
    b'_{L}) (\overline{d}'_{L}\gamma^\mu t_{L}) + 
    (\overline{u}_{L}\gamma_\mu t_{L}-\overline{d}'_{L}\gamma_\mu
    b'_{L}) (\overline{u}_{L}\gamma^\mu
    t_{L}-\overline{d}'_{L}\gamma^\mu b'_{L}),
  \end{aligned}
\end{equation}
comprise single-top and two-top vertices.  The ATLAS Coll.\
\cite{atl15ss} has obtained tight limits on the operators (\ref{eq:sst}),
through the term $(\overline{u}_L\gamma_\mu
t_L)(\overline{u}_L\gamma^\mu t_L)$, from its measurement of the
same-sign top production cross section (see section \ref{sec:contact}
below). 

We see from equation (23) of \cite{aguilaroperators} that
$O_{qq}^{(1)1133}$, $O_{qq}^{(3)3113}$, $O_{qq}^{(1)3113}$ enter the
decomposition into basis operators of the flavor-diagonal operators
$O^{11}_{qW}$ and $O^{33}_{qW}$, and both $O_{qq}^{(3)1113}$ and
$O_{qq}^{(3)3313}$ that of the flavor off-diagonal operator
$O_{qW}^{13}$. If we denote by $\mathsf{O}_{qq}$, $\mathsf{O}_{qq'}$
the four left-quark operators in the basis of \cite{aguilar4}, they
are related to those in (\ref{eq:4q.1}) by $\mathsf{O}_{qq}^{ijkl} =
1/2 O_{qq}^{(1)ijkl}$ and $\mathsf{O}_{qq'}^{ijkl}=1/4
(O_{qq}^{(3)ilkj}+O_{qq}^{(1)ilkj})$ \cite{aguilar4}.  We point out
also that the four-quark operator considered in \cite{fabbrichesi} is
$\hat{O}^{(3)}_{qq'}=O_{qq}^{(3)1133}$.

\section{Limits from decay processes}
\label{sec:fcnc}

In this section we discuss the limits on effective couplings that come
from several FCNC processes as well as those from observables
associated to $t\to Wq$ decays.  A global analysis of FCNC top-quark
interactions is given in \cite{maltoni15}, including NLO QCD
corrections \cite{zha14,zhu10}, in which many processess with direct
contributions from effective top vertices are surveyed to find those
yielding the best bounds on effective couplings.  Here, we restrict
ourselves to a simplified analysis involving only the two processes
that play the most important role in setting bounds for the operators
$O_{\varphi q}^{(-)k3}$, $O_{uZ}^{k3}$ and $O_{uA}^{k3}$: the on-shell
$t\to jZ$ decay and the single top $pp\to t\gamma ,\bar t \gamma$
production.  With these two experimental inputs we will be able to
obtain constraints similar to those in \cite{maltoni15}.  In addition,
we consider also two FCNC processes that are not associated to the top
but to the bottom quark: $B\to X_q \gamma$ and $Z\to b\bar q$.  These
will provide bounds on the operators $O_{dA}^{k3}$, $O_{dZ}^{k3}$ and
$O_{\varphi q}^{(+)k3}$.

\subsection{Limits from FCNC processes}
\label{sec:ncprocesses}

Let us briefly describe how we can obtain bounds for the NC part
of the operators.  We will start with the ones that do not
involve the top quark but the bottom quark.

%\subsection{Limits for  $O_{dA}^{3k}$ and $O_{uA}^{k3}$.}

The main contribution to the radiative decay $B\to X_q \gamma$
comes from the operator
$O_7=\frac{e}{16\pi^2} m_b \bar q_L \sigma^{\mu \nu} b_R F_{\mu \nu}$
with $q=d,s$, involving the right-handed $b$ quark and the
left-handed light quark $q$.  However, there is also a (smaller)
contribution from the right-handed $q$ operator $O_7^R$.
We observe that the operator $O^{3k}_{dA}$ directly (at tree level)
contributes to $O_7^R$ at the electroweak scale, with:
\bea
-\frac{4G_F}{\sqd}V_{tb} V^*_{tq} \frac{e}{16\pi^2} m_b C_7^R = 
 C^{3k}_{dA} y_t \frac{gs_W}{\sqd} \frac{v}{\Lambda^2}
\eea
In Ref.~\cite{tobias} (Eq. 42) we can find a specific expression
that singles out the contribution from $C^R_7$:
\bea
Br (B\to X_q \gamma) =
Br^{\rm SM} (B\to X_q \gamma) + 1.22 \times 10^{-2}
|V_{tq}|^2 \, \left|\frac{C^R_7}{C^{\rm SM}_7}\right|^2 
\eea
Where $C^{\rm SM}_7 (\mu = m_t) = -0.189$,
$V_{td}=0.0088$, $V_{ts}=0.0405$, and the SM values for
the branching fractions are given as \cite{tobias}:
\bea
10^4 Br^{\rm SM} (B\to X_s \gamma)
&=& 3.61 \pm 0.4 \, , \nonumber \\
10^5 Br^{\rm SM} (B\to X_d \gamma)
&=& 1.38 \pm 0.22 \, . \nonumber
\eea
We can use the experimental results \cite{expbqgamma}:
\bea
10^4 Br^{\rm exp} (B\to X_s \gamma)
&=& 3.43 \pm 0.21 \pm 0.07 \, , \nonumber \\
10^5 Br^{\rm exp} (B\to X_d \gamma)
&=& 0.92 \pm 0.30 \, , \nonumber
\eea
to set the limits $|C^{Rd}_7|<0.32$ and $|C^{Rs}_7|<0.36$ at 95\% CL.
When we translate these limits for the $C_{dA}$ coefficients we get an
extra suppression from the CKM matrix elements:
\bea
C^{31}_{dA} < 0.96 \times 10^{-5} \; ,
\;\;\;\;\;\;
C^{32}_{dA} < 5.4 \times 10^{-5} \; .
\label{bsglimits}
\eea
These are the strongest limits we have obtained for any of
the effective operators.  FCNC processes will also provide
the strongest constraints to all but the $O_{\varphi ud}$
operator as we shall see next.

%\subsection{Limits from $Z\to b\bar q$ decays}

Operators $O_{\varphi q}^{(+)k3}$ and $O_{dZ}$ with an
effective $Zbq$ coupling contribute directly to
$Z\to b\bar q$ decays ($q=d,s$).
We can use the ($90\%$C.L.) LEP upper limit \cite{lepzbq}
\bea
R_{bl} &=&
\frac{\Sigma_{q=d,s} \sigma (e^+e^- \to b\bar q , \bar b q)}
{\sigma (e^+e^- \to hadrons)} \;\; \leq \;
2.6 \times 10^{-3}
\eea
to set bounds on these coefficients.  Numerically, we can
write
\bea
\frac{\Gamma (Z\to b\bar q , \bar b q)}
{\Gamma (Z\to hadrons)} = \left(
2.63 |C_{\varphi q}^{(+)13}|^2 + 2.86 |C_{dZ}^{31}|^2 \right)
\times 10^{-3} \; + \; (13\to 23) \; ,
\eea
and obtain the following bounds
\bea
1.0 \left( |C_{\varphi q}^{(+)13}|^2 + |C_{\varphi q}^{(+)23}|^2
\right) \; + \; 1.1
\left( |C_{dZ}^{31}|^2 + |C_{dZ}^{32}|^2
\right) \leq 1.0 \label{zbqlimits0}
\eea

There are also (indirect) stringent bounds coming from the
$Br (B_d \to \mu^+ \mu^-)$ and
$Br (B_s \to \mu^+ \mu^-)$ measurements \cite{soreq}:
$|C_{\varphi q}^{(+)13}| < 0.005$ and $|C_{\varphi q}^{(+)23}| < 0.015$.

%\subsection{Limits from $t\to jZ$ decay}

The remaining operators that can be constrained with top-quark
FCNC processes are $O_{\varphi q}^{(-)k3}$, $O_{uZ}^{k3}$ and
$O_{uA}^{k3}$.  In Ref.~\cite{maltoni15} there is a thorough analysis
based on the $t\to jZ$ decay including off-shell contributions.
Let us simplify our discussion and consider the on-shell
$Br(t\to jZ)$ only:
\bea
Br(t\to jZ) = 3.34\times 10^{-4} \; \Sigma_{k=1,2} \;
\left( \left| \frac{C_{\varphi q}^{(-)k3}}{2x}-2xC^{k3}_{uZ}
\right|^2 + 2\left| 
\frac{C_{\varphi q}^{(-)k3}}{2}-2C^{k3}_{uZ} \right|^2 \right)
\nonumber
\eea
with $x=m_Z/m_t$ ($2x=1.05$). The ($95\%$CL) experimental upper
bound is: $Br(t\to jZ) < 5.0 \times 10^{-4}$ \cite{cmstjz},
therefore
\bea
\Sigma_{k=1,2} \;
\left( \left| \frac{C_{\varphi q}^{(-)k3}}{2x}-2xC^{k3}_{uZ}
\right|^2 + 2\left| 
\frac{C_{\varphi q}^{(-)k3}}{2}-2C^{k3}_{uZ} \right|^2 \right)
\; < \; 1.5  \label{tjzlimits}
\eea

For the other operator $O_{uA}^{k3}$ the CMS collaboration
has measured the process
$\sigma(pp\to t\gamma , \bar t \gamma)$
that provides the most stringent limit to date \cite{maltoni15}:
\bea
0.460 |C^{13}_{uA}|^2 + 0.037 |C^{23}_{uA}|^2 \, < \, 0.067
\label{pptglimits}
\eea

Equations (\ref{bsglimits}), (\ref{zbqlimits0}),
(\ref{tjzlimits}) and (\ref{pptglimits}) will be used to
define the allowed parameter regions for the $Wtq$
couplings.  They are based on the NC part of the dimension
six operators.  Below, we will describe the processes and
experimental values where the CC part plays the leading role.

\subsection{Limits from CC processes.}
\label{sec:ccprocesses}

We turn next to the charge-current decays $t\to Wq$, $q=d,s,b$.
Specifically, in this section we discuss the total width, the
branching ratios and $W$-helicity fractions in top decay.  From a
theoretical standpoint, it has been reported that the $t\to Wq$ decay
could get a $50\%$ enhancement in the context of the MSSM
\cite{lorenzowtd}, which underscores the importance of top decay
measurements like the ratio of $Br(t\to Wb)$ to $Br(t\to Wq)$
\cite{cmstwqratio}.  

In terms of form factors, the $t\to qW$ width for each
helicity of the $W$ boson, including terms proportional
to $m_q$, is given by
\cite{whelicity,whelicity2,aguilarwhel,czarnecki10}:  
\begin{subequations}
  \label{eq:helicities}
\begin{equation}
  \label{eq:helicities1}
  \begin{aligned}
\Gamma_0 &= A \left[ |a_t V^{q}_L-g^{q}_R|^2 + 
|a_t V^{q}_R-g^{q}_L|^2 + a_q G^{q}_0 \right],
 \\
\Gamma_+ &= A \left[ 2 |V^{q}_R - a_t g^{q}_L|^2 + a_q G^{q}_+
\right],
 \\
\Gamma_- &= A \left[ 2 |V^{q}_L -a_t g^{q}_R|^2 + a_q G^{q}_-
\right],
 \\
A &= \frac{g^2 m_t}{64\pi} \left( 1-\frac{m^2_W}{m^2_t} \right),
  \end{aligned}
\end{equation}
with
\begin{equation}
  \label{eq:helicities2}
  \begin{gathered}
G^{q}_0 = \left[ (a_t^2+1) (M^{q}_L+M^{q}_R)-2 a_t M^{q}_{LR}
\right]/(a_t^2-1),
 \quad
G^{q}_{+(-)} = M^{q}_{L(R)} - M^{q}_{R(L)} + G^{q}_0,
 \\
M^{q}_{LR} = 2 Re\{V^{q}_L V^{q*}_R + g^{q}_L g^{q*}_R\},
 \quad
M^{q}_{L(R)} = 2 Re\{V^{q}_{L(R)} g^{q*}_{L(R)}\},
  \end{gathered}
\end{equation}
\end{subequations}
where $a_t={m_t}/{m_W}$ and so is $a_q={m_q}/{m_W}$ for any down type
quark.  NLO QCD corrections (for $m_b=0$) to these W-helicity widths
can be found in \cite{drobnak10}.  We can use the expressions
(\ref{eq:helicities}) to obtain the ratio $Br(t\rightarrow Wb)/\sum
Br(t\rightarrow Wq)$, the total decay width, and the $W$-helicity
branching fractions.

The recent experimental measurement \cite{cmstwqratio} of the ratio:
\bea
{\cal R} &\equiv& \frac{Br(t\to Wb)}{\Sigma Br(t\to Wq)}
\; = \; 1.014 \; \pm 0.003\,(\textrm{stat}) \; \pm 0.032\,(\textrm{syst}) ,
\nonumber
\eea
is given by CMS also as a $95\%$ C.L.\ lower bound ${\cal R} > 0.955$ once
the condition ${\cal R} \leq 1$ has been imposed \cite{cmstwqratio}
(see also \cite{cdf13,d011} for previous Tevatron results).
We use this experimental lower bound on $\mathcal{R}$ to obtain the
following bounds at 95\% C.L.:
\begin{equation}
\label{twqbounds}
\left\{
\begin{aligned}
|C_{\varphi q}^{(\pm)k3}|,|C_{\varphi ud}^{3k}|  &< 7.29\\
|C_{uZ}^{k3}|,|C_{dZ}^{3k}| &< 3.16
\end{aligned}
\right.,
\;(k=1,2),\qquad\mathrm{or}\qquad
\left\{
\begin{aligned}
V^{q}_L,V^{q}_R &< 0.22\\
f^{q}_R,f^{q}_L &< 0.17
\end{aligned}
\right.,
\;(q=d,s),
\end{equation}
given here for convenience for both effective couplings and form
factors. 

Equation (\ref{eq:helicities}) also yields the $W$-helicity fractions
in $t\rightarrow bW$ decays:
\begin{equation}
  \label{eq:helicities3}
  F_0=\Gamma_0/\Gamma,
\qquad
  F_L=\Gamma_-/\Gamma.
\end{equation}
$F_{0,L}$ are the most sensitive observables to the flavor-diagonal
couplings $C^{33}_{\varphi ud}$, $C^{33}_{uZ}$, $C^{33}_{dZ}$, as has
long been known and duly exploited in the recent phenomenological
literature \cite{fabbrichesi,onofre}.  In this paper we take into
account the recent measurement of $W$-helicity fractions in top decays
from $t\overline{t}$ production at 8 TeV, with 20 fb$^{-1}$ of data
collected at the LHC \cite{cmswhelicity15}, which constitutes an
improvement from previous measurements 
\cite{cmswhelicitysingle}.  We obtain bounds for each effective
coupling at 95\% CL from the measured $W$-helicity fractions by means
of a likelihood analysis for the two correlated observables $F_{0,L}$,
as detailed in equations (18)--(20) of \cite{fabbrichesi}.  From the
CMS results \cite{cmswhelicity15},
\begin{equation}
  \label{eq:cmshelicityfracts}
  F_0=0.653\pm 0.016 (\mathrm{stat}) \pm 0.024 (\mathrm{syst}),
\quad
  F_L=0.329\pm 0.009 (\mathrm{stat}) \pm 0.025 (\mathrm{syst}),
\end{equation}
we get the single-coupling bounds:
\begin{equation}
  \label{eq:helbound}
\left\{
  \begin{gathered}
|C^{33}_{\varphi ud}| < 5.38,\quad
|C^{33}_{dZ}| < 1.27\\
-0.73 < C^{33}_{uZ\,r} < 1.63,\;
|C^{33}_{uZ\,i}| < 4.36
  \end{gathered}
\right.,
\qquad\mathrm{or}\qquad
\left\{
  \begin{gathered}
|V_R| < 0.16,\quad
|g_L| < 0.07\\
-0.09 < g_{R\,r} < 0.04,\;
|g_{R\,i}| < 0.24
  \end{gathered}
\right.
\end{equation}
The partial widths (\ref{eq:helicities1}) do not depend on the
left-handed vector couplings $C^{(\pm)33}_{\varphi q}$ (or,
equivalently, $V_L$) if the other
effective couplings vanish, so no single-coupling bounds are
obtained.  From the indirect measurement of the total top width
\cite{cmstwqratio} we obtain: 
\begin{equation*}
-3.02 < C^{(\pm)33}_{\varphi q\,r} < 3.7,\;
|\delta C^{(\pm)33}_{\varphi q\,i}|  < 16.13,
\qquad \mathrm{or} \qquad
-0.09 < \delta V_{L\,r} < 0.11,\;
|\delta V_{L\,i}|  < 0.48.
\end{equation*}
As discussed below, stronger bounds on $C^{(\pm)33}_{\varphi q}$ result from 
single-top production cross section.

\section{Single top quark production}
\label{sec:single.top.prod}

\begin{figure}[ht!]
\centering
%\fbox{
% tch.1.pdf  
\includegraphics[scale=0.7]{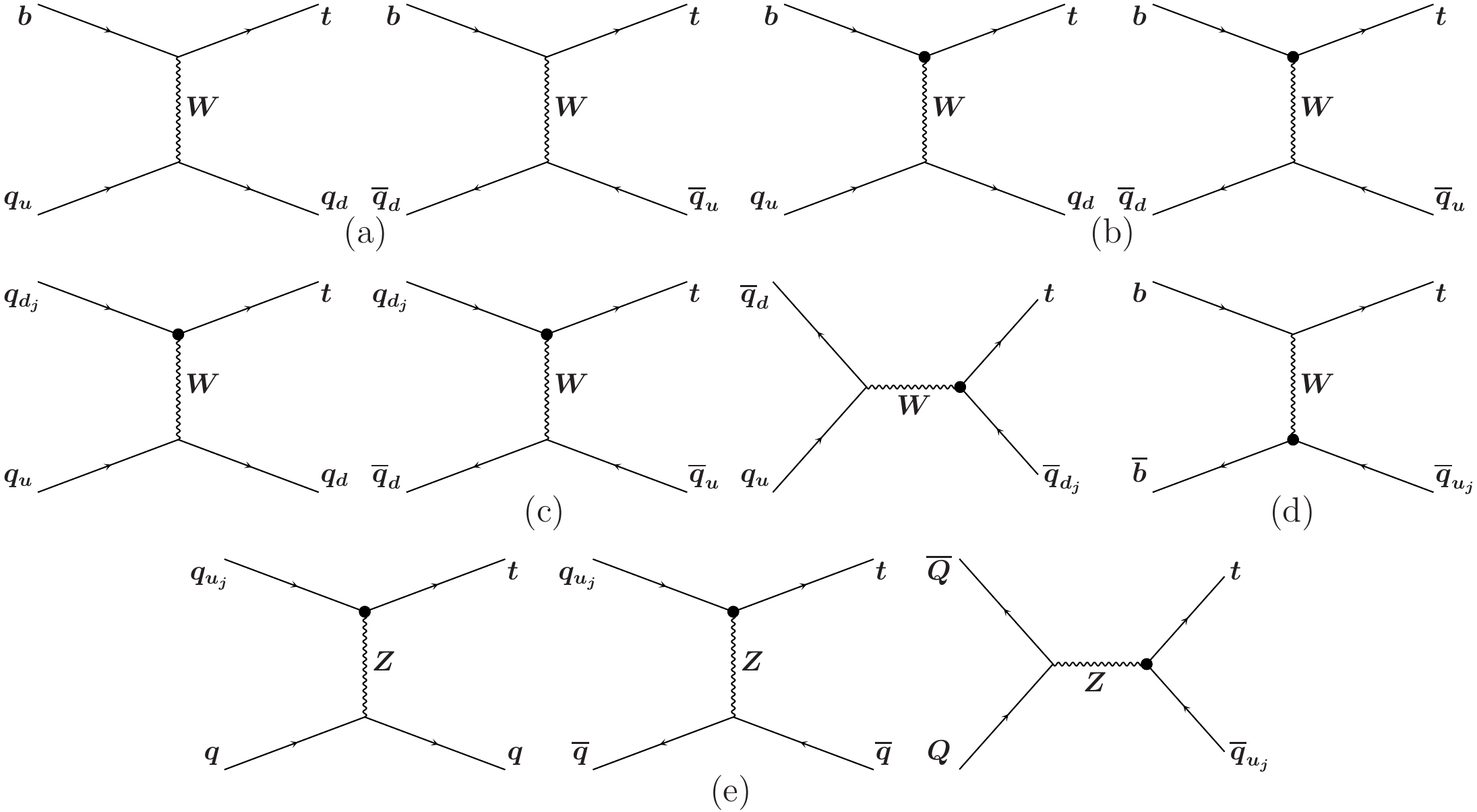}  
%}
\caption{Feynman diagrams for the process $pp\rightarrow tq$.  (a) SM
  diagrams, neglecting CKM mixing. (b) One flavor-diagonal CC
  effective vertex proportional to $C_{\varphi q}^{(\pm)33}$,
  $C_{\varphi ud}^{33}$, $C_{dZ}^{33}$ or $C_{uZ}^{33}$ ($q_u,q_d=u,d$
  or $c,s$).  (c) One flavor off-diagonal charged-current effective vertex proportional
  to $C_{\varphi q}^{(\pm)j3}$, $C_{\varphi ud}^{3j}$, $C_{dZ}^{3j}$
  or $C_{uZ}^{j3}$ ($j=1,2$). (d) One flavor off-diagonal charged-current effective
  vertex proportional to $C_{\varphi q}^{(\pm)j3}$ ($j=1,2$).  (e) One
  flavor-changing NC effective vertex proportional to $C_{\varphi
    q}^{(-)j3}$ or $C_{uZ}^{j3}$ ($j=1,2$; $q=u,d,c,s$; $Q=q,b$).}
  \label{fig:feyn1}
\end{figure}

\begin{figure}[ht!]
\centering
%\fbox{
% tch.2.pdf  
\includegraphics[scale=0.7]{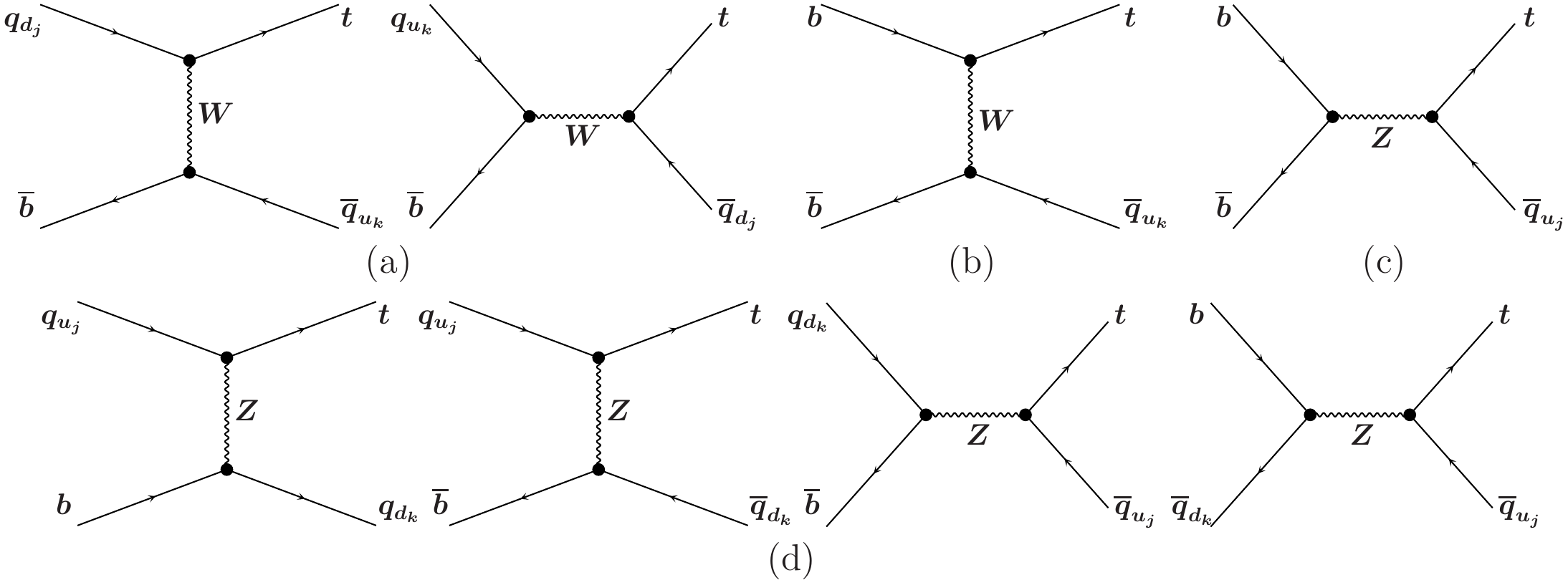}  
%}
\caption{Feynman diagrams for the process $pp\rightarrow tq$ with two
  effective vertices. (a) Two flavor off-diagonal charged-current vertices, one
  proportional to $C_{\varphi q}^{(\pm)k3}$ and one proportional to
  $C_{\varphi q}^{(\pm)j3}$, $C_{\varphi ud}^{3j}$, $C_{dZ}^{3j}$ or
  $C_{uZ}^{j3}$, $k,j=1,2$. (b) Two CC vertices, a flavor off-diagonal one
  proportional to $C_{\varphi q}^{(\pm)k3}$, $k=1,2$, and a
  flavor-diagonal one proportional to $C_{\varphi q}^{(\pm)33}$,
  $C_{\varphi ud}^{33}$, $C_{dZ}^{33}$ or $C_{uZ}^{33}$.  (c) Two
  NC vertices, a flavor-diagonal one proportional to
  $C_{\varphi q}^{(+)33}$ or $C_{dZ}^{33}$, and a flavor-changing one
  proportional to $C_{\varphi q}^{(-)j3}$ or $C_{uZ}^{j3}$,
  $j=1,2$. (d) Two flavor-changing NC vertices, one proportional to
  $C_{\varphi q}^{(+)k3}$ or $C_{dZ}^{3k}$ and one proportional to
  $C_{\varphi q}^{(-)j3}$ or $C_{uZ}^{j3}$, $k,j=1,2$.}
  \label{fig:feyn2}
\end{figure}

\begin{figure}[ht!]
\centering
%\fbox{
% sch.1.pdf
\includegraphics[scale=0.7]{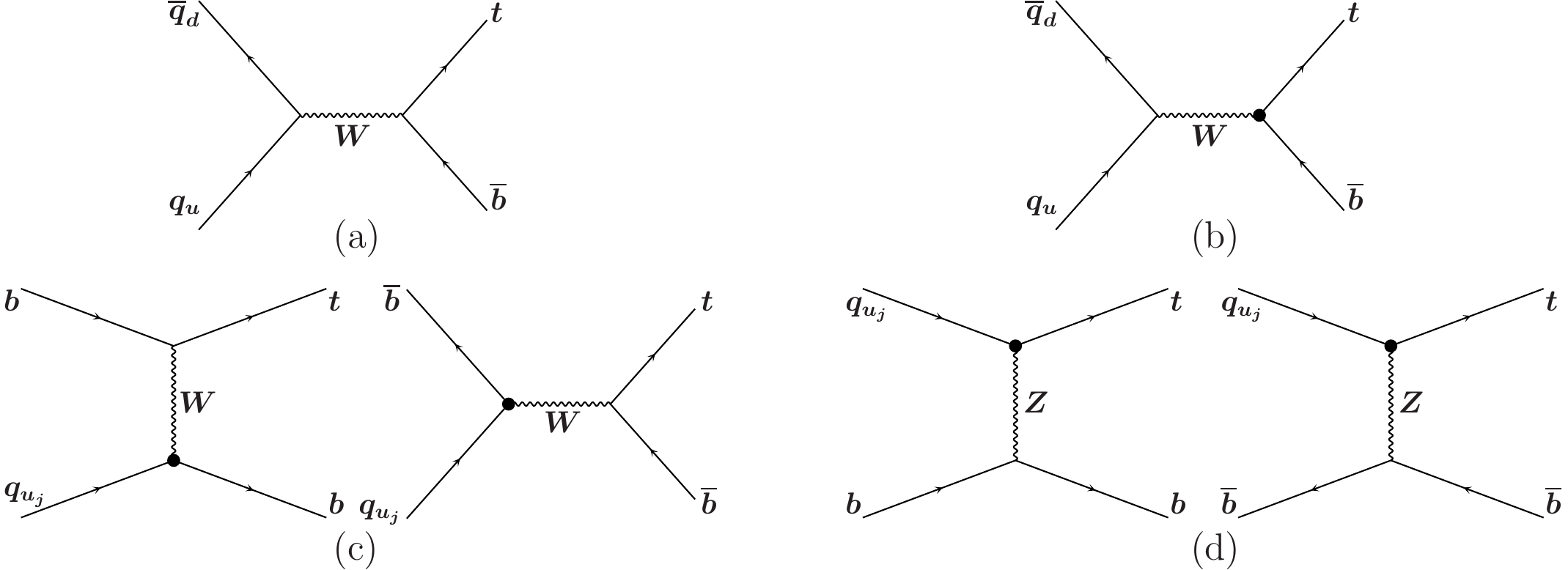}  
%}
\caption{Feynman diagrams for the process $pp\rightarrow
  tb,\,t\overline{b}$ with less than two effective vertices.  (a) SM
  diagram. (b) One flavor-diagonal CC effective vertex proportional to
  $C_{\varphi q}^{(\pm)33}$, $C_{\varphi ud}^{33}$, $C_{dZ}^{33}$ or
  $C_{uZ}^{33}$. (c) One flavor off-diagonal charged-current vertex effective proportional to
  $C_{\varphi q}^{(\pm)j3}$. (d) One FCNC effective vertex proportional to
  $C_{\varphi q}^{(-)j3}$ or $C_{uZ}^{(+)j3}$. ($j=1,2$.)}
  \label{fig:feyn3}
\end{figure}

\begin{figure}[ht!]
\centering
%\fbox{
% sch.2.pdf  
\includegraphics[scale=0.7]{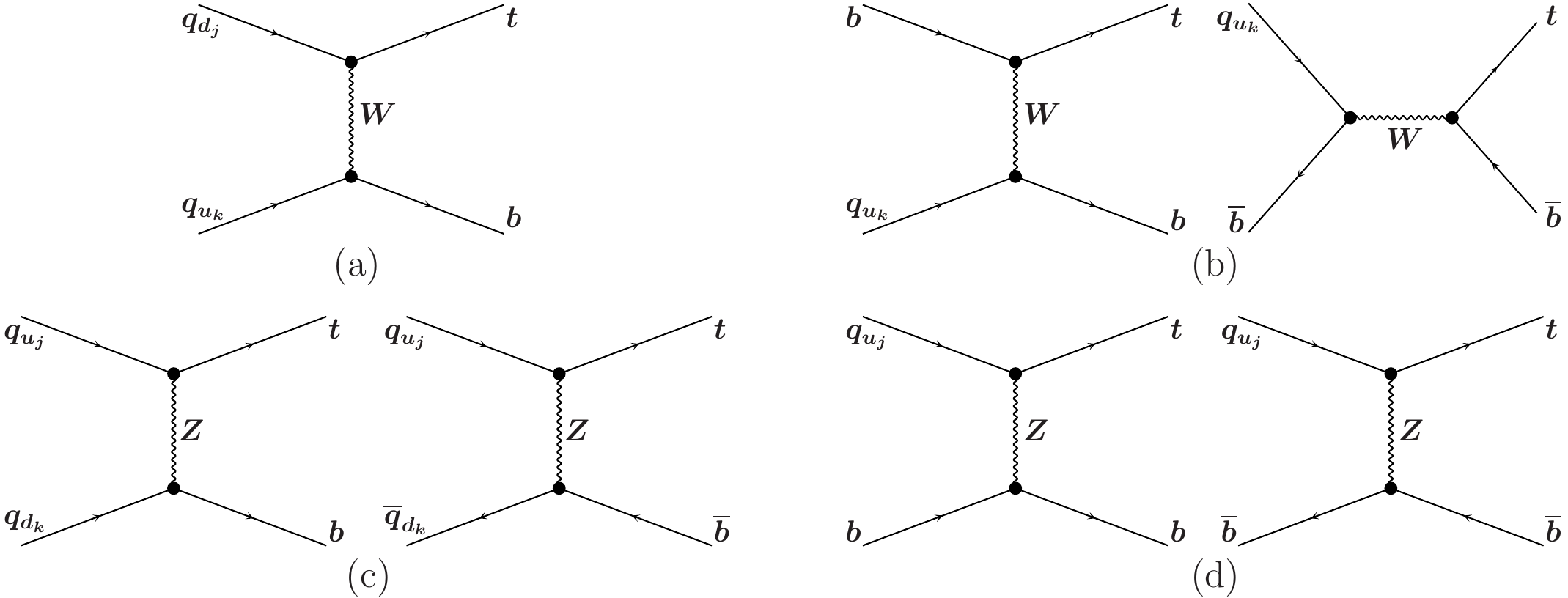}  
%}
\caption{Feynman diagrams for the process $pp\rightarrow
  tb,\,t\overline{b}$ with two effective vertices.  (a) Two
  flavor off-diagonal charged-current vertices, one proportional to $C_{\varphi
    q}^{(\pm)k3}$ and one to $C_{\varphi q}^{(\pm)j3}$, $C_{\varphi
    ud}^{3j}$, $C_{dZ}^{3j}$ or $C_{uZ}^{j3}$. (b) Two CC vertices, a
  flavor off-diagonal one proportional to $C_{\varphi q}^{(\pm)k3}$ and a
  flavor-diagonal one proportional to $C_{\varphi q}^{(\pm)33}$,
  $C_{\varphi ud}^{33}$, $C_{dZ}^{33}$ or $C_{uZ}^{33}$. (c) Two
  flavor-changing NC vertices, one proportional to $C_{\varphi
    q}^{(+)k3}$ or $C_{dZ}^{3k}$ and one to $C_{\varphi q}^{(-)k3}$ or
  $C_{uZ}^{j3}$. (d) Two NC vertices, a flavor-diagonal one
  proportional to $C_{\varphi q}^{(+)33}$ or $C_{dZ}^{33}$ and a
  flavor-changing one proportional to $C_{\varphi q}^{(-)j3}$ or
  $C_{uZ}^{j3}$. In all cases $k,j=1,2$.}
  \label{fig:feyn4}
\end{figure}

\begin{figure}[ht!]
\centering
%\fbox{
% tch.4q.pdf
\includegraphics[scale=0.7]{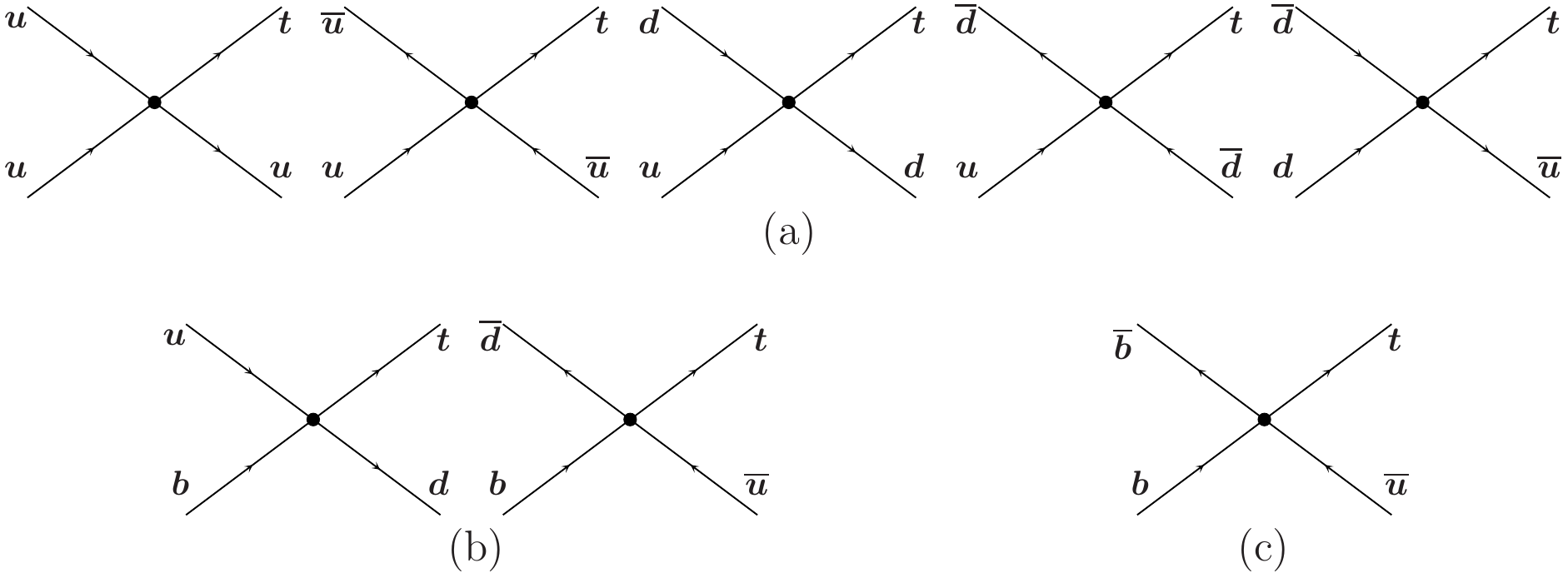}  
%}
\caption{Feynman diagrams for the process $pp\rightarrow tq$ with one
  contact-interaction four-quark vertex (a) proportional to
  $C_{qq}^{(1)1113}$ or $C_{qq}^{(3)1113}$, (b) proportional to
  $C_{qq}^{(1)3113}$ or $C_{qq}^{(3)1133}$, (c) proportional to
  $C_{qq}^{(1)3313}$ or $C_{qq}^{(3)3313}$.} 
  \label{fig:feyn5}
\end{figure}

\begin{figure}[ht!]
\centering
%\fbox{
% sch.4q.pdf
\includegraphics[scale=0.7]{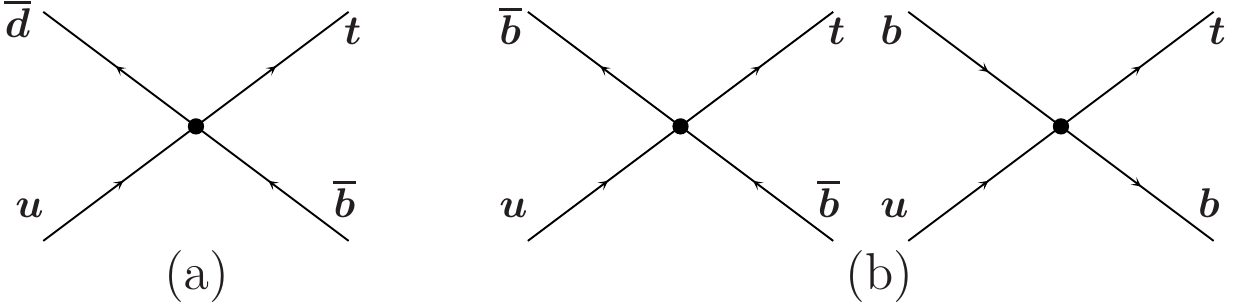}  
%}
\caption{Feynman diagrams for the process $pp\rightarrow
  tb,\,t\overline{b}$ with one contact-interaction four-quark vertex
  (a) proportional to $C_{qq}^{(1)3113}$ or $C_{qq}^{(3)1133}$, (b)
  proportional to $C_{qq}^{(1)3313}$ or $C_{qq}^{(3)3313}$.}
  \label{fig:feyn6}
\end{figure}

The effective operators (\ref{eq:operators}) contribute to the single
top production processes $pp\rightarrow tq$ (with $q$ a quark lighter
than $b$) and $p\overline{p}\rightarrow tb$, and to the associated
production process $p\overline{p}\rightarrow tW$, through both their CC
and NC vertices.  Furthermore, the four-quark operators
(\ref{eq:4q.2}) contribute to the first two types of single-top
production.  In this section we discuss single-top production assuming
for simplicity a diagonal CKM mixing matrix, to keep the diagrams down
to a manageable number.  Alternatively, the diagrams in Figures
\ref{fig:feyn1}--\ref{fig:feyn6} can be considered as given in the
weak-interaction quark basis.

The Feynman diagrams for the process $pp\rightarrow tq$ are shown in
Figures \ref{fig:feyn1} and \ref{fig:feyn2}.  In the SM, ignoring CKM
mixing, only the $tbW$ CC vertex can lead to single-top production in
$pp$ collisions, resulting in the four Feynman diagrams shown in
Figure \ref{fig:feyn1}(a).  Each one of the operators $O_{\varphi
  q}^{(\pm)33}$, $O_{\varphi ud}^{33}$, $O_{dZ}^{33}$ or $O_{uZ}^{33}$
contains a flavor-diagonal CC vertex, leading to four SM-like diagrams
with an effective $tbW$ vertex, Figure \ref{fig:feyn1}(b).  Flavor
off-diagonal charged-current effective vertices from the operators
$O_{\varphi q}^{(\pm)j3}$, $O_{\varphi ud}^{3j}$, $O_{dZ}^{3j}$ or
$O_{uZ}^{j3}$ lead to four $t$-channel and two $s$-channel diagrams
for each operator type and each value of $j=1,2$, Figure
\ref{fig:feyn1}(c).  The operators $O_{\varphi q}^{(\pm)j3}$ ($j=1,2$)
contain a flavor off-diagonal charged-current effective vertex for the
$b$ quark, giving rise to the diagram in Figure \ref{fig:feyn1}(d).
The flavor-changing NC effective vertices contained in $O_{\varphi
  q}^{(-)j3}$, $O_{uZ}^{j3}$ induce nine $t$/$u$-channel and five
$s$-channel diagrams mediated by a $Z$ boson for each operator type
and each value of $j=1,2$, Figure \ref{fig:feyn1}(e).

We point out here that the operator $O_{uW}^{j3}$ ($j=1,2$) would lead
to an additional set of diagrams analogous to those in Figure
\ref{fig:feyn1}(d), but mediated by a photon instead of a $Z$ boson.
The $\gamma$-mediated $t$-channel diagrams lead to a Coulomb
divergence in the full phase-space cross section, which is the reason
why we must consider the operators $O_{uZ}^{j3}$ instead of
$O_{uW}^{j3}$.  Notice that the extrapolation of detector-level
experimental data to a parton-level cross section for $pp\rightarrow
tq$ in full phase space, as given in \cite{cms2012,cms2014,atl14},
involves the explicit assumption of validity of the SM in which
flavor-changing photon vertices are absent.

The operators $O_{\varphi q}^{(\pm)k3}$ ($k=1,2,3$) contain flavor
diagonal and off-diagonal CC vertices involving a $b$ quark that
induce diagrams with two effective CC vertices, as shown in Figures
\ref{fig:feyn2} (a) and (b).  Furthermore, $O_{\varphi q}^{(+)n3}$ and
$O_{dW}^{3n}$ ($n=1,2,3$) contain flavor diagonal and off-diagonal NC
vertices involving a $b$ quark which, combined with the
flavor-changing NC vertices involving $t$ in $O_{\varphi q}^{(-)j3}$
and $O_{uZ}^{j3}$ ($j=1,2$) lead to the $Z$-mediated diagrams with two
effective vertices shown in Figures \ref{fig:feyn2} (c) and (d).  We
take into account in our analysis these diagrams with two effective
vertices, to check that their contributions are indeed small within
the allowed regions in coupling space, as discussed below.

The process $p\overline{p}\rightarrow tb$ also involves contributions
from both flavor off-diagonal and diagonal $tWq$ vertices.  The
associated Feynman diagrams involving one and two effective vertices
are displayed in Figures \ref{fig:feyn3} and \ref{fig:feyn4}, for
which a description completely analogous to the one given for the two
previous figures applies.  As for the associated $tW$ production, it
turns out not to play a relevant role in our results so we do not
dwell further on it here for brevity.

In Figure \ref{fig:feyn5} we show the Feynman diagrams for
$pp\rightarrow tq$ arising from the four-quark vertices from the
operators (\ref{eq:4q.2}).  As seen in the figure, in principle all
three types of operators (involving three, two and one light quark,
respectively) in (\ref{eq:4q.2}) contribute to this process.  It is
apparent from Figure \ref{fig:feyn5} (c), however, that the sensitiviy
of $tq$ production to operators with a single light quark must be
negligibly small due to the small PDF of the $b$ quark.  The
contribution of the four-quark operators with two and one light quarks
to $p\overline{p}\rightarrow tb$, $t\overline{b}$ production is shown
in Figure \ref{fig:feyn6}.  It is in connection with these diagrams
that $tb$ production plays its most important role in this paper,
since it furnishes the only available limits on the four-quark
couplings (\ref{eq:4q.2}) with a single light quark and the tightest
ones on those with two light quarks.  In this section we have
restricted our discussion of four-quark operators to those involving
only first- and third-generation quarks for brevity.  However, the
extension of equation (\ref{eq:4q.2}) and diagrams \ref{fig:feyn5},
\ref{fig:feyn6} to include second-generation quarks is
straightforward.  In section \ref{sec:pheno} below we discuss, besides
the operators (\ref{eq:4q.2}), also those four-quark couplings
involving second-generation quarks to which single-top production
possesses significant sensitivity.

In our computations of single-top production cross sections we always
take into account the decay vertex $t\rightarrow bW$, not shown in
Figures \ref{fig:feyn1}--\ref{fig:feyn4} for simplicity, which can
proceed through the SM vertex or flavor-diagonal effective ones. This
leads to the cross section $\sigma(pp\rightarrow tq \rightarrow
bWq)=\sigma(pp\rightarrow tq) Br(t \rightarrow bW)$, with $Br(t
\rightarrow bW)$ the branching fraction for this decay mode.  If we
restrict ourselves to flavor-diagonal effective operators the decay
vertex is irrelevant, since $Br(t \rightarrow bW)$ cannot depend on
flavor-diagonal couplings and therefore it cancels in the ratio
$\sigma_\mathrm{eff}/\sigma_\mathrm{SM}$ of the effective and SM cross
sections.  When, as in this study, flavor off-diagonal vertices are
considered, the branching fraction cannot be ignored since it does
depend on those couplings.  We find that the dependence of $Br(t
\rightarrow bW)$ on the off-diagonal effective couplings tends to
relax the bounds on those couplings relative to the ones that would
be obtained from the pure production cross section, without including
top decay, by up to 15\% for operators involving first-generation
quarks.

The effective cross section for single-top production can be expressed
perturbatively as a power series in the effective couplings.  As seen
from Figures \ref{fig:feyn1}--\ref{fig:feyn4}, the cross section for
production and decay receives contributions from the effective
vertices up to the sixth power in the dim-6 effective couplings.
Higher powers arise from the additional dependence of the top
propagator on effective couplings.  We have explicitly verified in all
the cases discussed below that, for values of the effective couplings
within their allowed regions, the effect of terms with powers higher
than quadratic is negligibly small.

Due to the GIM mechanism for FCNC and to the smallness of CKM
third-generation mixing in flavor off-diagonal charged-currents,
flavor off-diagonal processes involving the top quark are strongly
suppressed at tree level (and beyond) in the SM.  For that reason,
terms linear in flavor off-diagonal dim-6 effective couplings in the cross
section ($\mathcal{O}(1/\Lambda^2)$) are negligibly small since they
arise from the interference of amplitudes involving a dim-6 effective
vertex with the SM amplitude.  By the same token, the contributions to
the cross section at order $1/\Lambda^4$ of flavor off-diagonal dim-8
operators are also suppressed.  At that order, however, there can be
contributions from dim-8 flavor-diagonal operators interfering with
the SM which, although expected to be small, are currently unknown and
constitute an inherent uncertainty of the EFT analysis.

On the other hand, that uncertainty does not affect the
flavor-diagonal couplings $C_{\varphi q}^{(\pm)33}$ and $C_{uZ}^{33}$,
which contribute to the single-top production cross section dominantly
through linear terms at order $1/\Lambda^2$ from interference with the
SM model.  The other two flavor-diagonal couplings, $C_{\varphi
  ud}^{33}$ and $C_{dW}^{33}$, have their linear interference terms
suppressed by $m_b$ and, therefore, significantly smaller.

\subsection{Statistical analysis}
\label{sec:statanal}

Let the experimental and theoretical SM cross sections for single top
production in $pp$ or $p\overline{p}$ collisions be
\begin{equation}
  \label{eq:xsctexpthr}
{\sigma_\mathrm{exp}}\!\!
\begin{array}{l}
+\;\Delta{\sigma_\mathrm{exp}^\uparrow}\\
-\;\Delta\sigma_\mathrm{exp}^\downarrow  
\end{array}
=\sigma_\mathrm{exp}\times\left( 1\!\!
\begin{array}{l}
+\;\varepsilon_\mathrm{exp}^\uparrow\\  
-\;\varepsilon_\mathrm{exp}^\downarrow
\end{array}
 \right),
\qquad 
\sigma_\mathrm{thr}\!\!
\begin{array}{l}
+\;\Delta{\sigma_\mathrm{thr}^\uparrow}\\
-\;\Delta\sigma_\mathrm{thr}^\downarrow  
\end{array}
=\sigma_\mathrm{thr}\times\left( 1\!\!
\begin{array}{l}
+\;\varepsilon_\mathrm{thr}^\uparrow\\  
-\;\varepsilon_\mathrm{thr}^\downarrow
\end{array}
 \right),
\end{equation}
where we have allowed for asymmetrical uncertainties.  The theoretical
cross section $\sigma_\mathrm{thr}$ is assumed to be computed in the
SM, possibly at NNLO+NNLL (e.g., \cite{kidonakis2,kid10,kid10sch}).
We denote by $\widetilde{\sigma}(\lambda)$ the cross section in the
effective theory, computed at LO in the effective couplings $\lambda$
and at the same order as $\sigma_\mathrm{thr}$ in the SM couplings, so
that $\widetilde{\sigma}(0)=\sigma_\mathrm{thr}$.  Furthermore, we
denote by $\sigma(\lambda)$ the cross section in the effective theory
computed at LO in both the effective couplings and the SM, and
$K(\lambda)=\widetilde{\sigma}(\lambda)/\sigma(\lambda)$, so that
$K(0)$ is the $K$-factor in the SM.  We base our analysis on the
inequalities
\begin{equation}
  \label{eq:ineq.1}
\widetilde{\sigma}(\lambda) - \widetilde{\sigma}(0)  
\lessgtr \sigma_\mathrm{exp}-\sigma_\mathrm{thr}\!\!
\begin{array}{l}
  +\; \sqrt{\sigma_\mathrm{exp}^2\varepsilon_\mathrm{exp}^{\uparrow\,2}
  + \sigma_\mathrm{thr}^2\varepsilon_\mathrm{thr}^{\downarrow\,2}}\\
  -\; \sqrt{\sigma_\mathrm{exp}^2\varepsilon_\mathrm{exp}^{\downarrow\,2}
  + \sigma_\mathrm{thr}^2\varepsilon_\mathrm{thr}^{\uparrow\,2}}
\end{array}.
\end{equation}
Dividing both sides by $\sigma_\mathrm{thr}$ we get,
\begin{equation}
  \label{eq:ineq.2}
\frac{K(\lambda)}{K(0)} \frac{\sigma(\lambda)}{\sigma(0)}
\lessgtr R_\mathrm{exp} \left(1\!\!
\begin{array}{l}
  +\; \sqrt{\varepsilon_\mathrm{exp}^{\uparrow\,2}
  + \varepsilon_\mathrm{thr}^{\downarrow\,2}/R_\mathrm{exp}^2}\\
  -\; \sqrt{\varepsilon_\mathrm{exp}^{\downarrow\,2}
  + \varepsilon_\mathrm{thr}^{\uparrow\,2}/R_\mathrm{exp}^2}
\end{array}  \right),
\end{equation}
with $R_\mathrm{exp}=\sigma_\mathrm{exp}/\sigma_\mathrm{thr}$.  In
(\ref{eq:ineq.2}) the factor
$K(\lambda)/K(0)=1+\mathcal{O}(\alpha_s \lambda)$, 
so at LO in the SM we set it to 1. Thus, finally, at LO in both the
effective and the SM couplings, we get
\begin{equation}
  \label{eq:ineq.3}
\frac{\sigma(\lambda)}{\sigma(0)}
\lessgtr R_\mathrm{exp} \left(1\!\!
\begin{array}{l}
  +\; \sqrt{\varepsilon_\mathrm{exp}^{\uparrow\,2}
  + \varepsilon_\mathrm{thr}^{\downarrow\,2}/R_\mathrm{exp}^2}\\
  -\; \sqrt{\varepsilon_\mathrm{exp}^{\downarrow\,2}
  + \varepsilon_\mathrm{thr}^{\uparrow\,2}/R_\mathrm{exp}^2}
\end{array}  \right).
\end{equation}
On the right-hand side we identify the ratio $R_\mathrm{exp}$ of cross
sections for single-top production and decay $t\rightarrow bW$ with
the ratio of production cross sections, from which it differs by
multiplication of both numerator and denominator by $Br(t\rightarrow
bW) = |V_{tb}|^2 \pm \mathcal{O}(10^{-4})$.  On the left-hand side of
(\ref{eq:ineq.3}) the LO cross section $\sigma(\lambda)$ enters only
through the ratio $R(\lambda)=\sigma(\lambda)/\sigma(0)$, which does
not depend on the tree-level cross section normalization.
Furthermore, for small values of $\lambda$, the relative scale and PDF
uncertainties are much smaller for $R(\lambda)$ than for the cross
sections themselves.

We point out, parenthetically, that (\ref{eq:ineq.1}) is different
from the similarly-looking equation
\begin{equation}
  \label{eq:ineq.4}
\sigma(\lambda) - \sigma(0)  
\lessgtr \sigma_\mathrm{exp}-\sigma_\mathrm{thr}\!\!
\begin{array}{l}
  +\; \sqrt{\sigma_\mathrm{exp}^2\varepsilon_\mathrm{exp}^{\uparrow\,2}
  + \sigma_\mathrm{thr}^2\varepsilon_\mathrm{thr}^{\downarrow\,2}}\\
  -\; \sqrt{\sigma_\mathrm{exp}^2\varepsilon_\mathrm{exp}^{\downarrow\,2}
  + \sigma_\mathrm{thr}^2\varepsilon_\mathrm{thr}^{\uparrow\,2}}
\end{array}.
\end{equation}
This inequality does depend on the tree-level cross section
normalization.  Equation (\ref{eq:ineq.4}) can be rewritten as
\begin{equation}
  \label{eq:ineq.5}
\frac{\sigma(\lambda)}{\sigma(0)}
\lessgtr K(0) R_\mathrm{exp} \left(1\!\!
\begin{array}{l}
  +\; \sqrt{\varepsilon_\mathrm{exp}^{\uparrow\,2}
  + \varepsilon_\mathrm{thr}^{\downarrow\,2}/R_\mathrm{exp}^2}\\
  -\; \sqrt{\varepsilon_\mathrm{exp}^{\downarrow\,2}
  + \varepsilon_\mathrm{thr}^{\uparrow\,2}/R_\mathrm{exp}^2}
\end{array}  \right) + 1 - K(0),
\end{equation}
which is different from (\ref{eq:ineq.3}), in particular, because it
depends explicitly on $K(0)$, and therefore also on the normalization
of the tree-level cross section. In the case of the single-top
production cross sections for combined $tq+\overline{t}q$ production
measured at the LHC, from the SM results of \cite{kidonakis2} and our
tree-level results we get $K(0)=1.07$.  As a result, the bounds on
effective couplings determined by (\ref{eq:ineq.3}) are only slightly
tighter than those obtained from (\ref{eq:ineq.5}).  On the other
hand, for $tb$ production at the Tevatron, from the SM result of
\cite{kid10sch} we get $K(0)=1.67$, which leads to significantly more
restrictive bounds obtained from (\ref{eq:ineq.3}) than from
(\ref{eq:ineq.5}).

\section{Results}
\label{sec:pheno}

In this section we present the results obtained from the processes
considered in sections \ref{sec:fcnc} and \ref{sec:single.top.prod}, as 
single-coupling limits and as two-coupling allowed regions for $Wtq$,
$Wtb$ and four-quark effective interactions.  Our results are based on
the cross sections for $tq$ production measured by CMS:
\begin{equation}
  \label{eq:cmstq}
\begin{tabular}{ccccc}
    $\sigma(pp\rightarrow tq+\overline{t}q) = (67.2\pm6.1)$ pb,
    & 7 TeV, 2.73 fb$^{-1}$ &\cite{cms2012},\\[2pt]  
    $
      \left\{
      \begin{aligned}
    \sigma(pp\rightarrow tq+\overline{t}q) &= (83.6\pm7.75)
    \mathrm{pb}\\
    \sigma(pp\rightarrow tq) &= (53.8\pm4.65) \mathrm{pb}\\
    \sigma(pp\rightarrow \overline{t}q) &= (27.6\pm3.92) \mathrm{pb}
      \end{aligned}
      \right.,
    $
    & 8 TeV, 19.7 fb$^{-1}$ &\cite{cms2014},
\end{tabular}
\end{equation}
together with the NNLO SM predictions from \cite{kidonakis2}
\begin{equation}
  \label{eq:thrtq}
\begin{tabular}{cc}
    $\sigma(pp\rightarrow tq+\overline{t}q) = (64.6\!\!
\begin{array}{l}
+2.1\\
-0.6  
\end{array} \!\!
\begin{array}{l}
+1.5\\
-1.7  
\end{array})$pb,
    & 7 TeV, \\[2pt]  
    $
      \left\{
      \begin{aligned}
    \sigma(pp\rightarrow tq+\overline{t}q) &= (87.2\!\!
\begin{array}{l}
+2.8\\
-1.0  
\end{array}\!\!
\begin{array}{l}
+2.0\\
-2.2  
\end{array})
    \mathrm{pb}\\
    \sigma(pp\rightarrow tq) &= (56.4\pm\!\!
\begin{array}{l}
+2.1\\
-0.3  
\end{array}\pm 1.1) \mathrm{pb}\\
    \sigma(pp\rightarrow \overline{t}q) &= (30.7\pm0.7\!\!
\begin{array}{l}
+0.9\\
-1.1  
\end{array}) \mathrm{pb}
      \end{aligned}
      \right.,
    $
    & 8 TeV.  
\end{tabular}
\end{equation}
Further results on the $tq$ production cross section at 7 TeV at the
LHC, and on differential cross sections,  have been given by ATLAS
\cite{atl14}.  We comment on those data below in section
\ref{sec:diffxsct}.   The cross section for $tb$ production has been
measured by CDF and D0:
\begin{equation}
  \label{eq:cdftb}
\begin{tabular}{ccccc}
    $\sigma(p\overline{p}\rightarrow t\overline{b}+\overline{t}b) = (1.29\!\!
\begin{array}{l}
+0.26\\
-0.24  
\end{array})$ pb,
    & 1.96 TeV, 19.4 fb$^{-1}$ &\cite{cdf14a},
\end{tabular}
\end{equation}
and computed at NNLO in the SM in \cite{kid10sch}:
\begin{equation}
  \label{eq:thrtb}
\begin{tabular}{cc}
    $\sigma(pp\rightarrow t\overline{b}+\overline{t}b) = (1.05\pm 0.06)$pb,
    & 1.96 TeV.
\end{tabular}
\end{equation}
Notice that $tb$ production has also been observed at the LHC
\cite{atl15,cms13b}.  We have also taken into account the associated
$tW$ production cross section measured by CMS \cite{cms13a,cms14b},
for which an approximate NNLO SM result is given in \cite{kidtw}.  We
include as well in our results the measurements of $W$-helicity
fractions in top decays, top decay width and ratio of branching
fractions $Br(t\rightarrow Wb)/\sum Br(t\rightarrow Wq)$ discussed in
section \ref{sec:ccprocesses}, as well as the various decay processes
in section \ref{sec:ncprocesses}.

Some four-quark operators receive bounds from the $t\overline{t}$
production cross section (see section \ref{sec:contact} below).  In
those cases we use the experimental measurements:
\begin{equation}
  \label{eq:ttxxsct}
\begin{tabular}{rrr}
    $\sigma(pp\rightarrow t\overline{t}) = (239\pm 12.7)$ pb,
    & 8 TeV, 5.3 fb$^{-1}$ &\cite{cmsttx13},\\
    $\sigma(pp\rightarrow t\overline{t}) = (158.1\pm 11)$ pb,
    & 7 TeV, 2.3 fb$^{-1}$ &\cite{cmsttx12b},\\    
    $\sigma(p\overline{p}\rightarrow t\overline{t}) = (7.6\pm 0.41)$ pb,
    & 1.96 TeV, 8.8 fb$^{-1}$ &\cite{cdfd0ttx14},
\end{tabular}
\end{equation}
together with the NNLO SM results:
\begin{equation}
  \label{eq:ttxxsctthr}
\begin{tabular}{rrr}
    $\sigma(pp\rightarrow t\overline{t}) = (252.9\!\!
\begin{array}{l}
+13.3\\
-14.5  
\end{array})$ pb,
    & 8 TeV,  &\cite{cza11,cza13},\\
    $\sigma(pp\rightarrow t\overline{t}) = (163\!\!
\begin{array}{l}
+11.4\\
-10.3  
\end{array})$ pb,
    & 7 TeV,  &\cite{kid10},\\
    $\sigma(p\overline{p}\rightarrow t\overline{t}) = (7.35\!\!
\begin{array}{l}
+0.28\\
-0.33  
\end{array})$ pb,
    & 1.96 TeV,  &\cite{cza11,cza13},
\end{tabular}
\end{equation}
as quoted by the experimental collaborations.  NNLO SM results for
$t\overline{t}$ production at 7 TeV have also been given in
\cite{ahr10,ali11}, which are consistent with the values quoted
above. 

We compute the tree-level cross sections for single-top production
and decay with the matrix-element Monte Carlo program
\textsc{MadGraph5\_aMC@NLO} version 2.2.3 \cite{mg5,mg5d}.  The
effective operators were implemented in \textsc{MadGraph5} by means of
the UFO \cite{ufo11} interface of the program \textsc{FeynRules}
version 2.0.33 \cite{feynrul}.  In all cases we set $m_t=172.5$ GeV,
$m_b=4.7$ GeV, $m_Z=91.1735$ GeV, $m_W=80.401$ GeV, $m_h=125$ GeV,
$\alpha(m_Z)=1/132.507$, $G_F=1.1664\times10^{-5}$ GeV$^{-2}$,
$\alpha_S(m_Z)=0.118$, and the Higgs vacuum-expectation value
$v=246.22$ GeV.  We set the renormalization and factorization
scales fixed at $\mu_R=m_t=\mu_F$ and use the parton-distribution
functions CTEQ6--L1 as implemented in \textsc{MadGraph5}.  The new
physics scale $\Lambda$ is set to 1 TeV.  Furthermore, we take into
account full CKM mixing in our computations, though its effects on our
results are very limited.  As expected, third-generation mixing is
negligible and could be safely ignored.  For values of the effective
couplings within the allowed regions obtained here, Cabibbo mixing
becomes relevant only for certain four-quark operators, as discussed
in more detail below. 

\subsection{Limits on flavor off-diagonal couplings}
\label{sec:flavor.changing}

In Table~\ref{fcnclimits} we gather 95\% CL limits on flavor
off-diagonal $Wtq$ effective couplings taken to be non-zero one at a
time.  All operators in (\ref{eq:operators}) involve both $W$ and
$Z/A$ bosons, except for $O^{3k}_{\varphi ud}$. Thus, in the table we
show limits originating from processes involving vertices $Vtq$ with
$V$ a charged or neutral vector boson, and with $q$ a first-generation
quark (upper two rows) or second-generation one (lower two rows).  On
the first row we give the best bounds on those couplings involving
flavor off-diagonal charged-current vertices $Wtd$, obtained from CMS
data for single $t$ production at 8 TeV \cite{cms2014}.  The cross
section for combined $t+\overline{t}$ production at the same energy
leads to somewhat weaker bounds, as seen in the figures below.  The
tightest limits for effective couplings associated to the flavor
off-diagonal charged-current vertices $Wts$ stem from the ratio of top
branching fractions $Br(t\rightarrow Wb)/\sum_qBr(t\rightarrow Wq)$
\cite{cmstwqratio}, and are shown on the third row of the table.  We
remark that direct bounds on $C^{3k}_{\varphi ud}$, $k=1,2$, have not
been given in the previous literature.  However, an indirect limit
$|C^{31}_{\varphi ud}|<5\times10^{-3}$ is given in \cite{cri11} based
on the contribution of $O^{31}_{\varphi ud}$ to $b\to d\gamma$.

On the second and fourth rows of Table~\ref{fcnclimits} we display the
bounds on those same couplings obtained from FCNC processess.  For the
bounds on operators $O_{\varphi q}^{(+)k3}$ and $O_{dZ}^{3k}$ we have
used the decay $Z\to b \bar d(\bar s)$ in Eq.~(\ref{zbqlimits0}).
Operators $O_{dA}^{3k}$ contribute directly to $Br (B\to X_q \gamma)$
and for them we obtain the strongest bounds in this study as seen in
Eq.~(\ref{bsglimits}).  For the bounds on operators $O_{\varphi
  q}^{(-)k3}$ and $O_{uZ}^{3k}$ we have used the decay $t\to Zu(c)$
(with on-shell $Z$) in Eq.~(\ref{tjzlimits}) 
\cite{maltoni15}.  Finally, the best bounds on $O_{uA}^{3k}$ come from
the FCNC single top production process $\sigma(pp\to t\gamma , \bar t
\gamma)$ \cite{maltoni15}.
\begin{table}
\begin{center}
  \begin{tabular}{c c|c|c|c|c|c|c|c|}\cline{3-9}
& & \rule[-10pt]{0pt}{25pt}$\left|C^{3k}_{\varphi ud}\right|$ & $\left|C^{(-)k3}_{\varphi
    q}\right|$ & $\left|C^{k3}_{uZ}\right|$ &
$\left|C^{k3}_{uA}\right|$  & $\left|C^{(+)k3}_{\varphi q}\right|$ &
$\left|C^{3k}_{dZ}\right|$ & $\left|C^{3k}_{dA}\right|$ \\\hline\hline 
\multicolumn{1}{|c|}{\raisebox{-16pt}[0pt][0pt]{\rotatebox{90}{$k=1$}}}
&\rule[-4pt]{0.pt}{15pt} $Wtd$    & $5.30$ & $2.68$ & $1.37$ & --
& $4.95$ & $1.96$ &  --                 \\\cline{2-9} 
\multicolumn{1}{|c|}{}
&\rule[-4pt]{0.pt}{15pt} $Z(A)tu$ & --     & $1.09$ & $0.42$ & $0.38$
& $1.00$ & $0.95$ & $0.96\times 10^{-5}$ \\\hline 
\multicolumn{1}{|c|}{\raisebox{-17pt}[0pt][0pt]{\rotatebox{90}{$k=2$}}}
&\rule[-4pt]{0.pt}{15pt} $Wts$    & $7.29$ & $7.29$ & $3.16$ & --
& $7.29$ & $3.16$ &  --                 \\\cline{2-9} 
\multicolumn{1}{|c|}{}
&\rule[-4pt]{0.pt}{15pt} $Z(A)tc$ & --     & $1.08$ & $0.42$ & $1.35$
& $1.00$ & $0.95$ & $5.4\times 10^{-5}$ \\\hline \hline 
\multicolumn{2}{c}{} & 
\multicolumn{1}{|c|}{no NC} & 
\multicolumn{3}{c|}{up-quark FCNC} & 
\multicolumn{3}{c|}{down-quark FCNC}  \\\cline{3-9}
  \end{tabular}
\end{center}
\caption{Limits on the seven operators that are relevant for
  the study on $Wtq$ couplings.  A comparison is made between
  limits coming from processes involving flavor off-diagonal
  charged-current interactions (first and third rows) and purely FCNC
  processes (second and fourth rows).}  
\label{fcnclimits}
\end{table}

\begin{figure}[ht!]
  \centering
% fgr.im.pd.pdf
\includegraphics[scale=0.88]{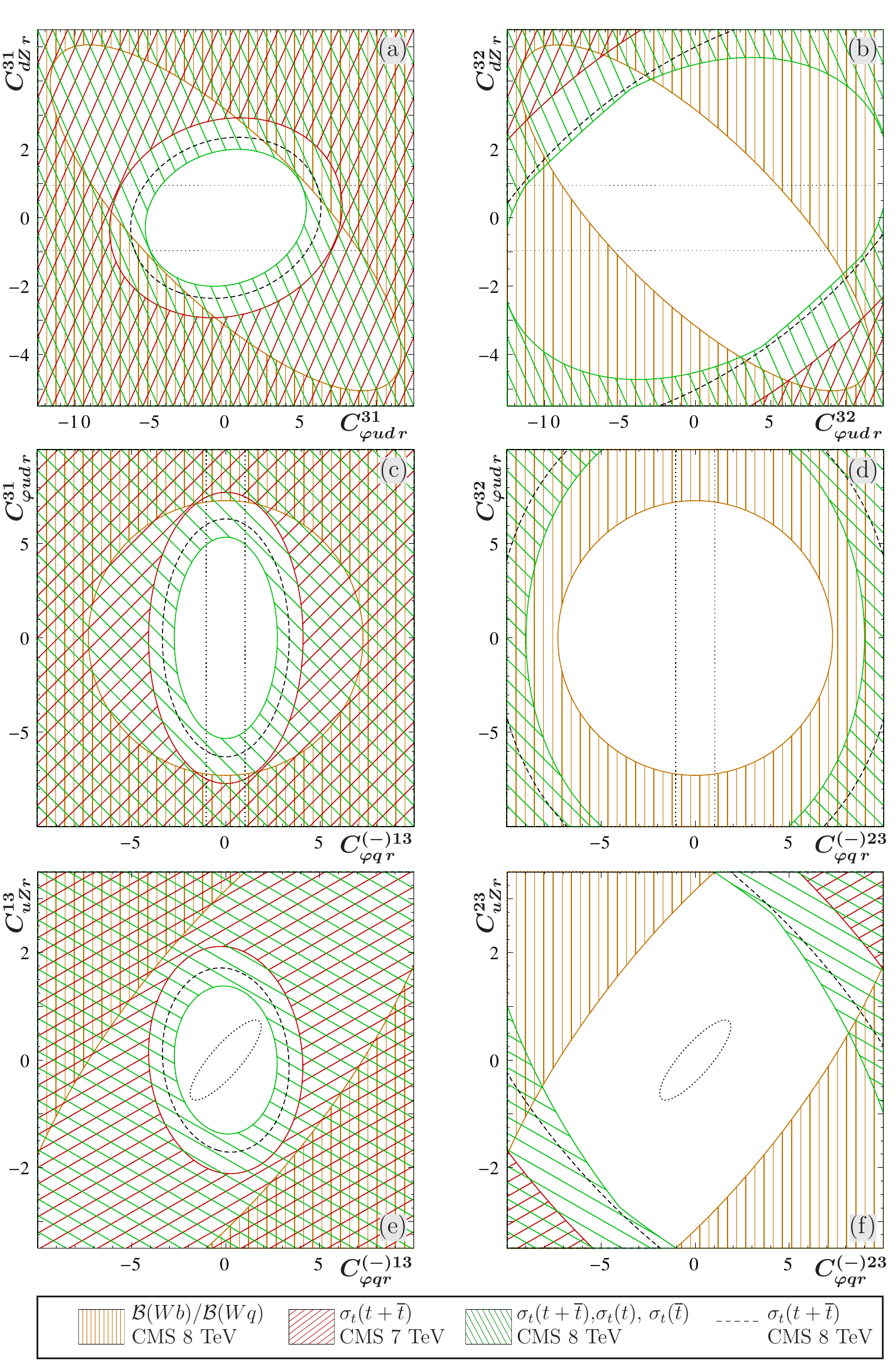}  
  \caption{Parameter regions for flavor off-diagonal $Wtq$ effective
    couplings allowed at 95\% CL. 
    Orange hatched area: region excluded by the branching fractions
    $Br(Wb)/\sum_iBr(Wq_i)$ \cite{cmstwqratio} in
    top decays. Red hatched area: region excluded by the cross section
    for $pp\rightarrow tq+\overline{t}q$ at 7 TeV \cite{cms2012}.
    Green hatched area: region excluded by the cross sections for
    $pp\rightarrow tq+\overline{t}q$, $tq$, $\overline{t}q$ at 8 TeV
    \cite{cms2014}. Black dashed line: region excluded by the cross
    section for $pp\rightarrow tq+\overline{t}q$ at 8 TeV alone
    \cite{cms2014}. Dotted lines: (a) and (b), bounds on
    $|C_{dZ}^{3j}|$ ($j=1,2$) from (\ref{zbqlimits0});
    (c)--(f), allowed regions from (\ref{tjzlimits}).
}
  \label{fig:offdiagonal2}
\end{figure}

Besides the single-coupling bounds in Table \ref{fcnclimits} we
consider also several allowed two-parameter regions.  In Figure
\ref{fig:offdiagonal2} we display allowed regions for pairs of
effective couplings having non-vanishing interference
($C^{3k}_{\varphi ud}$/$C^{3k}_{dZ}$ and $C^{(-)k3}_{\varphi
  ud}$/$C^{k3}_{uZ}$, $k=1,2$) and for vector couplings
($C^{3k}_{\varphi ud}$/$C^{(-)k3}_{\varphi q}$). Those regions are
obtained at 95\% CL, as described in section \ref{sec:statanal}, from
the production cross section for $tq+\overline{t}q$, with $q$ lighter
than $b$, in $pp$ collisions at 7 TeV \cite{cms2012} (red hatched area
in the figure), from the production cross section for
$tq+\overline{t}q$ at 8 TeV \cite{cms2014} (black dashed line), from
the intersection of the regions allowed by the production cross
sections for $tq$, $\overline{t}q$ and $tq+\overline{t}q$ at 8 TeV
\cite{cms2014} (green hatched area), and from the ratio of branching
fractions $Br(t\rightarrow Wb)/\sum_qBr(t\rightarrow Wq)$ in $t$ decay
\cite{cmstwqratio} (orange hatched area).  Also shown in the figure,
for comparison, are the bounds on $C_{dZ}^{3k}$ and
$C^{(-)k3}_{\varphi ud}$ from FCNC processes as given in Table
\ref{fcnclimits} (black dotted lines in Figure \ref{fig:offdiagonal2}
(a)--(d)), and the allowed region for $C^{(-)k3}_{\varphi
  q}$/$C^{k3}_{uZ}$ from the branching fraction $Br(t\rightarrow jZ)$
as given by (\ref{tjzlimits}) (black dotted lines in Figure
\ref{fig:offdiagonal2} (e)--(f)).

The cross section for $tb$ production has been measured at the
Tevatron \cite{cdf14a,cdf14b} and at the LHC \cite{cms13b,atl15}. The
production process (see Figures \ref{fig:feyn3}, \ref{fig:feyn4}) does
not depend on $C^{3k}_{\varphi ud}$, $C^{3k}_{dZ}$ and has a modest
sensitivity to $C^{(\pm)k3}_{\varphi ud}$, $C^{k3}_{uZ}$.  In fact,
for $tb$ production followed by $t\rightarrow Wb$ decay, most of the
sensitivity to the effective couplings originates in the dependence on
them of the branching fraction $Br(t\rightarrow Wb)$, which
is already explicitly taken into account in Figure
\ref{fig:offdiagonal2}. For this reason we do not include $tb$
production in this figure.  We have also taken into account $tW$
associated production, whose cross section has been measured at the
LHC at 7 and 8 TeV \cite{cms13a,cms14b}.  Due to the somewhat large
current experimental uncertainties in those measurements (30\% at 7
and 23\% at 8 TeV), the allowed regions resulting from this process
are significantly looser than those shown in the figure, so we omit
them for the sake of simplicity.

\begin{figure}[ht!]
  \centering
% fgr.fgh.pd.pdf
\includegraphics[scale=0.85]{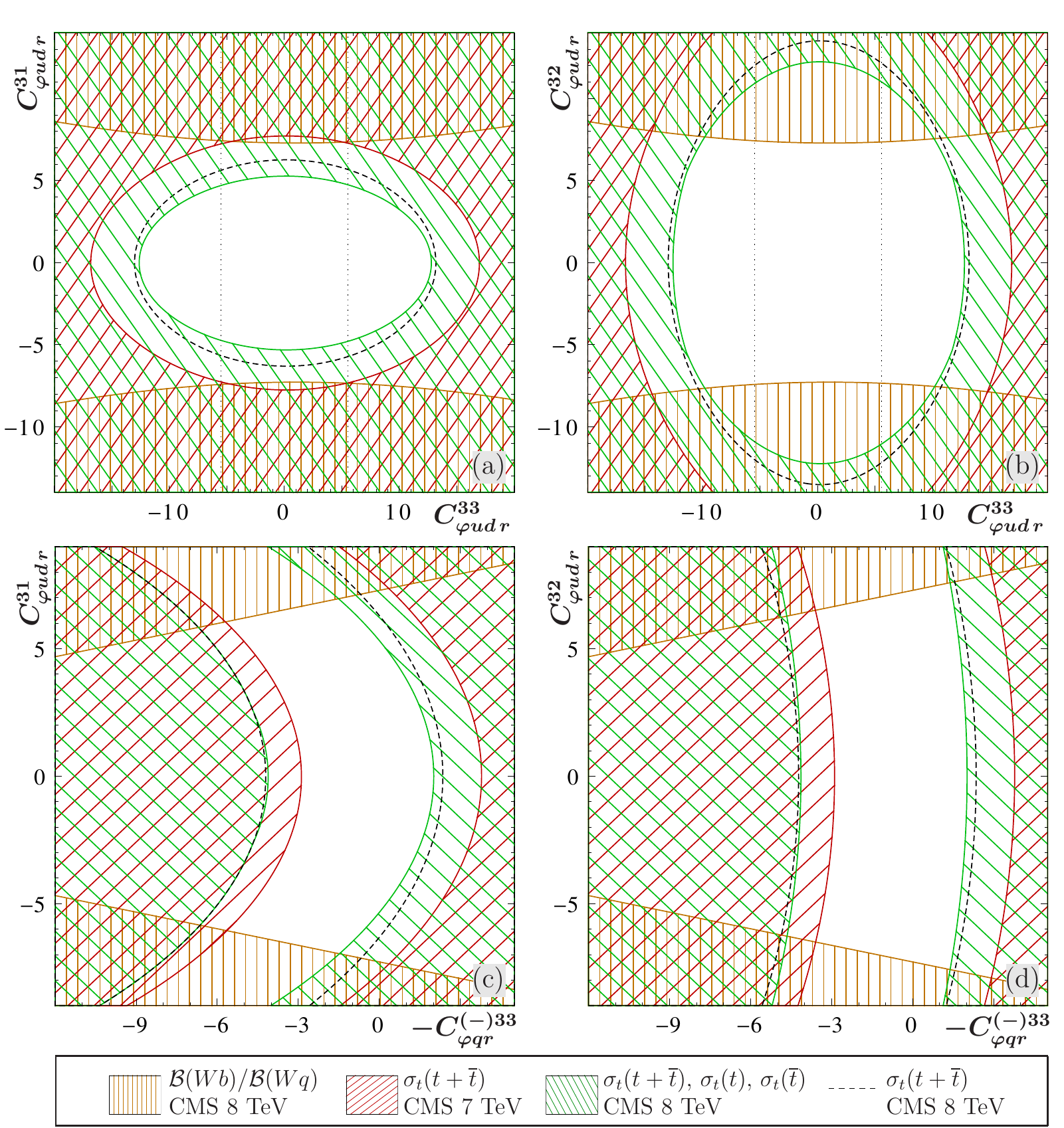}  
  \caption{Allowed parameter regions at 95\% CL for the flavor
    off-diagonal right-handed vector $Wtq$ effective couplings $C_{\varphi ud\,
      r}^{3j}$ ($j=1,2$) versus flavor-diagonal left- and
    right-handed vector ones.  Color codes as in the previous figure.
    Black dotted lines in (a) and (b): bounds on $C_{\varphi ud\,
      r}^{33}$ from $W$-helicity fractions in top decays
    \cite{cmswhelicity15}. 
  }
  \label{fig:diag-offdiag}
\end{figure}

In Figure \ref{fig:diag-offdiag} we show the allowed regions on the
plane of the flavor off-diagonal charged-current right-handed vector
couplings $C^{3k}_{\varphi ud}$, $k=1,2$, and the flavor diagonal
left- and right-handed vector couplings $C^{33}_{\varphi ud}$ and
$C^{(-)33}_{\varphi q}$. The allowed regions are determined by the
same experimental data as used in the previous figure.  In Figure
\ref{fig:diag-offdiag} (a), (b) we include for reference the bounds on
$C^{33}_{\varphi ud}$ set by the experimental determination of $W$
helicity fractions in $t$ decays
\cite{cmswhelicity15,cmswhelicitysingle} (black dotted lines in the
figure), as discussed in more detail below.  In the case of the
flavor-diagonal coupling $C^{(-)33}_{\varphi q}$, the best bounds
result from a combination of single-top production cross sections at 7
and 8 TeV \cite{cms2012,cms2014} as seen in the figure.  We remark
that for the processes used in the figure the operators
$O^{(+)33}_{\varphi q}$ and $-O^{(-)33}_{\varphi q}$ are equivalent,
so the the coupling $C^{(+)33}_{\varphi q}$ can be used equally well
instead of $-C^{(-)33}_{\varphi q}$ to label the horizontal axes in
Figures \ref{fig:diag-offdiag} (c), (d).

%\clearpage 

%\clearpage

\subsection{Limits on the $Wtb$ couplings}

\begin{figure}[ht!]
%\fbox{
% fgr.7de.pd.pdf
\includegraphics[scale=0.88]{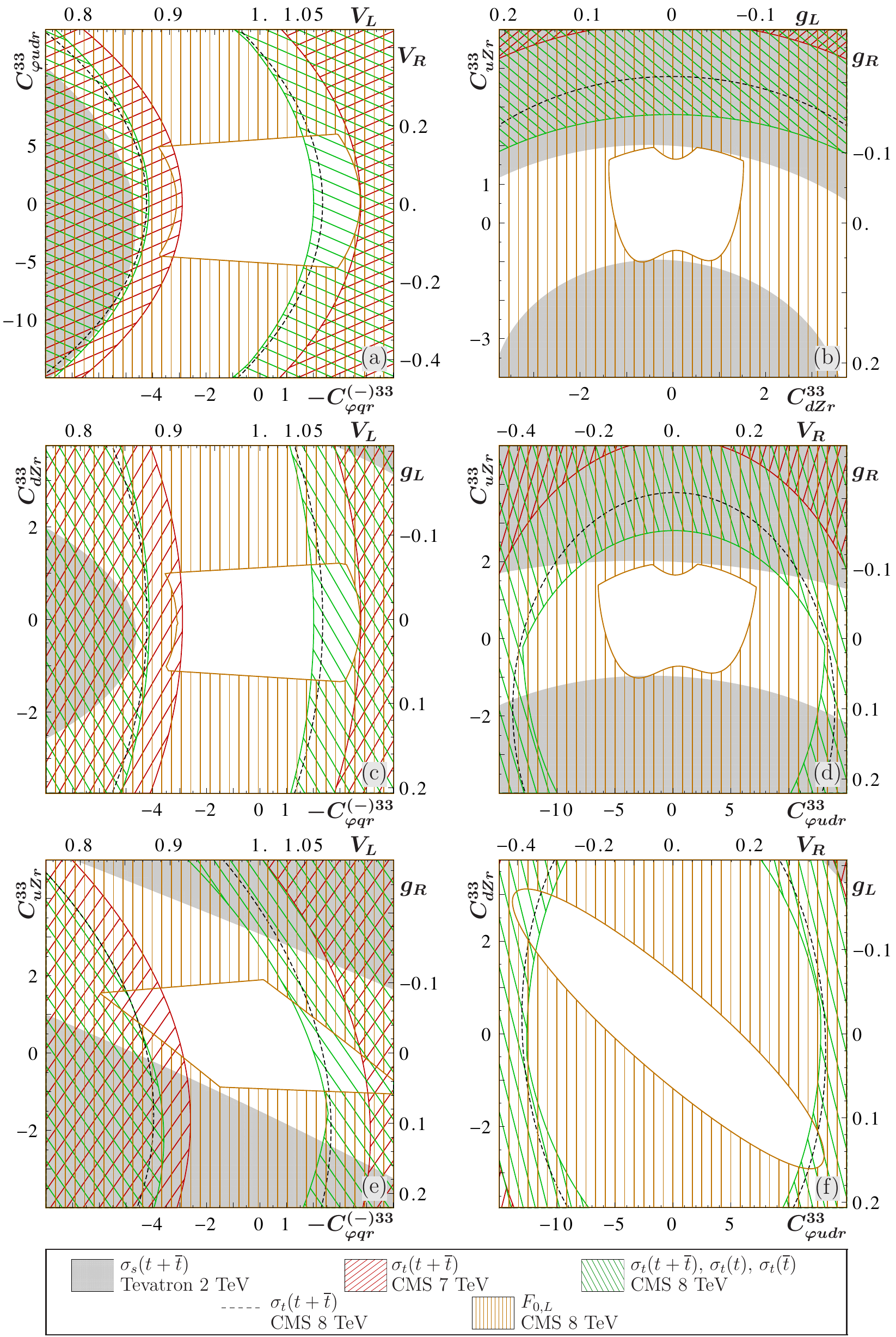}  
%}
  \caption{Allowed parameter regions at 95\% CL for flavor-diagonal
$Wtq$ effective couplings.  
Orange hatched area: region excluded by $W$-helicity fractions 
in top decays \cite{cmswhelicity15} and top decay width \cite{cmstwqratio}. Gray area: region excluded by the
cross section for $p\overline{p}\rightarrow
t\overline{b}+\overline{t}b$ at 1.96 TeV \cite{cdf14a}.  Red and
green hatched areas and dashed line as in Figure
\ref{fig:offdiagonal2}.  
%(fig:7de)
}
  \label{fig:7de}
\end{figure}

In Figure \ref{fig:7de} we show allowed regions for all possible pairs
of $Wtb$ effective couplings.  As noticed above in connection with
Figure \ref{fig:diag-offdiag}, in this context the coupling
$C^{(+)33}_{\varphi q}$ is equivalent to $-C^{(-)33}_{\varphi q}$.  We
obtain the bounds in this figure from the same set of cross sections
for single-top production together with a light jet at the LHC
\cite{cms2012,cms2014} as in Figure \ref{fig:offdiagonal2}. Also shown
in this figure (as light-gray areas) are the allowed regions resulting
from the cross section for $p\overline{p}\rightarrow
t\overline{b}+\overline{t}b$ measured at the Tevatron \cite{cdf14a}
(see also \cite{cdf14b} for related Tevatron results, and
\cite{atl15,cms13b} for measurements at the LHC).  As seen in Figure
\ref{fig:7de}, the most restrictive limits on the couplings
$C^{33}_{\varphi ud}$, $C^{33}_{uZ}$, $C^{33}_{dZ}$ are imposed by the
combination of $W$-helicity fractions and decay width (orange hatched
area).  On the other hand, $F_{0,L}$ have a weak dependence on
$C^{(\pm)33}_{\varphi q}$, which is bounded by the decay width and
single-top production cross sections.  The best bounds on
$C^{(\pm)33}_{\varphi q}$ are set by the $tq$ production cross
sections at 7 TeV (lower bound) and at 8 TeV (upper bound).  The
intersection of the regions allowed by $tq$, $\overline{t}q$ and
$tq+\overline{t}q$ production at 8 TeV (green hatched areas) is
necessarily more restrictive than the region obtained from
$tq+\overline{t}q$ production alone, the difference between the two
being most apparent for $C^{33}_{uZ}$ and less pronounced in the case
of $C^{(\pm)33}_{\varphi q}$.

%\clearpage

\subsection{Differential cross sections}
\label{sec:diffxsct}

Besides the total cross sections for single-top production used in the
previous sections, we have considered also the total and differential
cross sections reported by ATLAS for separate and combined single top
and antitop production in the LHC at 7 TeV with a total integrated
luminosity of 4.58 fb$^{-1}$.

\begin{figure}[ht!]
  \centering
% fgr.cms.atl.pdf
\includegraphics[scale=0.85]{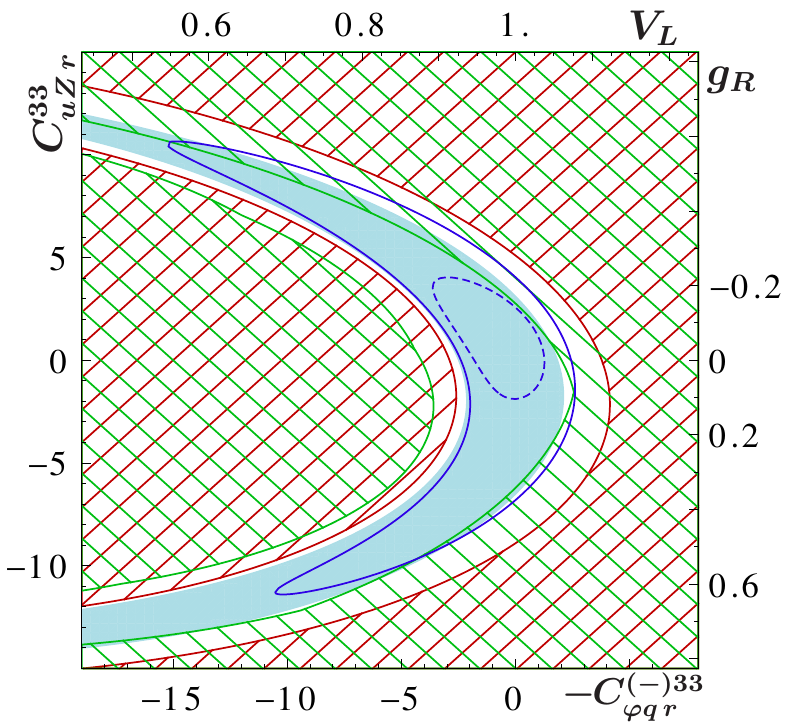}  
  \caption{Allowed parameter regions at 95\% CL for two
flavor-diagonal  $Wtq$ effective couplings.  Red and
green hatched areas as in previous figures.  Light-blue area: allowed
region at 95\% CL determined simultaneously by the total cross sections for
$pp\rightarrow tq+\overline{t}q$ at 7 TeV \cite{cms2012}, for
$pp\rightarrow tq+\overline{t}q$, $tq$, $\overline{t}q$ at 7 TeV
\cite{atl14}, and for $pp\rightarrow tq+\overline{t}q$, $tq$,
$\overline{t}q$ at 8 TeV \cite{cms2014}.  Dark-blue solid line: allowed
region at 95\% CL determined simultaneously by the total cross
sections and the differential cross sections $d\sigma/d|y|(t)$,
$d\sigma/d|y|(\overline{t})$, $d\sigma/d|\vec{p}_T|(t)$,
$d\sigma/d|\vec{p}_T|(\overline{t})$, excluding the two
highest-$|\vec{p}_T|$ bins. Dark-blue dashed line: allowed
region at 95\% CL determined simultaneously by total and
differential cross sections, including all bins.
}
  \label{fig:cms.atl}
\end{figure}

The effect of these additional data on the allowed parameter regions
is illustrated in Figure \ref{fig:cms.atl} for the couplings
$C_{\varphi q\,r}^{(-)33}/C_{uZ\,r}^{33}$.  For reference, we include
in Figure \ref{fig:cms.atl} the same 95\% CL--allowed regions as in
Figure \ref{fig:7de} (e).  We combined those CMS cross sections with
the total cross sections for $pp\rightarrow tq+\overline{t}q$, $tq$,
$\overline{t}q$ at 7 TeV measured by ATLAS \cite{atl14} in a $\chi^2$
analysis, to obtain at 95\% CL the allowed region shown by the
light-blue band in Figure \ref{fig:cms.atl}.  As seen in the figure,
the allowed region is little changed in a neighborhood of the origin
by the inclusion of the additional data points.

We further extended the analysis by including all bins in the measured
differential cross sections $d\sigma/d|y|(t)$,
$d\sigma/d|y|(\overline{t})$ in the $\chi^2$ function, with their
correlation matrices, as well as the data for
$d\sigma/d|\vec{p}_T|(t)$, $d\sigma/d|\vec{p}_T|(\overline{t})$
excluding the two highest-$|\vec{p}_T|$ bins in each distribution
(four excluded bins in total).  The resulting allowed region at 95\%
CL is shown in Figure \ref{fig:cms.atl} by the blue solid line.
Finally, adding the previously excluded highest-$|\vec{p}_T|$ bins to
the $\chi^2$ function, yields the allowed region delimited by the blue
dashed line in the figure.  As seen there, those highest-$|\vec{p}_T|$
bins have a large effect on the allowed region, which we attribute to
the fact that their central values show large deviations
($\sim2\sigma$) from the SM NLO predictions, especially in the case of
$d\sigma/d|\vec{p}_T|(\overline{t})$.  We point out as well that the
highest-$|\vec{p}_T|$ bin corresponds to an energy range of
$150-500$GeV that is relatively close to the new physics scale
$\Lambda =1$ TeV assumed here, which would make the validity of our obtained
bounds uncertain.

We conclude that the results from the LHC Run-I (7 TeV) on single-top
production do not significantly add to the constraining we have
obtained based on Run-II (8TeV) data.  This is true even after
considering the input from the absolute rapidity and $p_T$
distributions, unless we take into consideration the large deviations
observed in the last two bins.

%%%%%%%%%%%%%%%%%%%%%%%%%%%%%%%%%%%%%%%%%%%%%%%%%%%%%%%

\subsection{Limits on four-fermion operators}
\label{sec:contact}

The operators $O_{qq}^{(1,3)ijk3}$ with $i,j,k<3$ contribute to
single-top production only through $pp\rightarrow tq,$
$t\overline{q}$.  The Feynman diagrams related to these vertices are
shown in Figure \ref{fig:feyn5} (a).  Due to their flavor off-diagonal
nature, there is no interference of these diagrams with the SM ones
due to the very small third-generation mixing.  Yet, there is an
enhanced sensitivity to these couplings because of the large
first-generation PDFs.  Taking only one coupling to be non-zero at a
time, from the single-top production cross section
$\sigma(pp\rightarrow tq)$ at 8 TeV \cite{cms2014} we get the
following single-coupling bounds.  The operators $O_{qq}^{(1,3)1113}$,
$O_{qq}^{(1,3)1213}$ receive the strongest bounds among four-quark
operators: 
\begin{equation}
  \label{eq:bnds.1113}
  |C_{qq}^{(1)1113}|,|C_{qq}^{(1)1213}| < 0.30, \qquad  
  |C_{qq}^{(3)1113}|,|C_{qq}^{(3)1213}| < 0.23.  
\end{equation}
The cross section for antitop production at 8 TeV, and the combined
$t+\overline{t}$ production cross sections at 8 and 7 TeV
\cite{cms2012,cms2014} lead to somewhat weaker bounds.  Similarly, for
the analogous four-quark operators involving only one third-generation
quark, we obtain the single-coupling bounds:
\begin{equation}
  \label{eq:bnds.13}
  |C_{qq}^{(1)1123}| < 1.23, \quad    
  |C_{qq}^{(3)1123}| < 0.50, \quad
  |C_{qq}^{(1)2113}| < 0.86, \quad
  |C_{qq}^{(3)2113}| < 0.72. 
\end{equation}

The operators $O_{qq}^{(1,3)3113}$ and $O_{qq}^{(3)1133}$ contribute
to both single-top production channels ($pp\rightarrow tq,$
$t\overline{q}$ and $pp\rightarrow t\overline{b},$ $tb$) and to
$t\overline{t}$ production, while $O_{qq}^{(1)1133}$ contributes only
to the latter process.  Notice that these four operators are
Hermitian, so their couplings are real.  The bounds we find are
\begin{equation}
  \label{eq:bnds.4}
  \begin{gathered}
  -1.07 < C_{qq}^{(1)3113} < 1.19, \quad      
  -0.80 < C_{qq}^{(3)3113} < 0.96,\\
  -2.94 < C_{qq}^{(1)1133} < 2.67, \quad      
  -0.18 < C_{qq}^{(3)1133} < 0.36.
  \end{gathered}
\end{equation}
The bounds on $C_{qq}^{(1)3113}$ result from a combination of the ones
obtained from $t\overline{t}$ production ($-1.07 < C_{qq}^{(1)3113} <
1.23$) and those from $tb$ production ($-2.19 < C_{qq}^{(1)3113} <
1.19$), both at the Tevatron.  As mentioned in section \ref{sec:4q},
$O_{qq}^{(3)3113}=-O_{qq}^{(1)3113}+$ terms with 0 or 2 top fields, so
single-top production does not distinguish between the two.  The
bounds on $C_{qq}^{(3)3113}$ in (\ref{eq:bnds.4}) arise from
$t\overline{t}$ production at the Tevatron.  The operator
$O_{qq}^{(1)1133}$ does not contribute to single-top production.  The
limits (\ref{eq:bnds.4}) on $C_{qq}^{(1)1133}$ are a combination of
the ones obtained from $t\overline{t}$ production, at 8 TeV at the LHC
($-2.94 < C_{qq}^{(1)1133} < 2.80$) and at the Tevatron ($-3.28 <
C_{qq}^{(1)1133} < 2.67$).  The tightest limits on $C_{qq}^{(3)1133}$
arise from $tb$ production at the Tevatron, the bounds from
$t\overline{t}$ production on that coupling being much looser, $\sim3$
at 8 TeV and larger at lower energies.  As discussed in section
\ref{sec:4q}, the operator $O_{qq}^{1313}$ contributes to single-top
and $tt$ production and $O_{qq}^{3131}=O_{qq}^{1313\,\dagger}$ to
single-top and $\overline{t}\overline{t}$ production.  Bounds on
$C_{qq}^{(1,3)1313}$ have been given by ATLAS \cite{atl15ss} from their
measurement of same-sign $tt$ production at 8 TeV.  We quote here the
ATLAS result for completeness, which in our conventions reads
$|C_{qq\,r}^{(1)1313}|,|C_{qq\,r}^{(3)1313}|< 0.0265$ at 95\% CL.  

The sensitivity of both the $tq$ and $tb$ production processes to the
couplings $C_{qq}^{(1)3123}$, $C_{qq}^{(3)1233}$ is significantly
enhanced by Cabibbo mixing.  The strongest bounds on those couplings
arise from $tq$ production at 8 TeV:
\begin{equation}
  \label{eq:bnds.4a}
  \begin{aligned}
    -2.25 < &C_{qq\,r}^{(1)3123} < 2.22,\quad
    -2.23 < &C_{qq\,i}^{(1)3123} < 2.23,\\
    -1.14 < &C_{qq\,r}^{(3)1233} < 1.09, \quad
    -1.11 < &C_{qq\,i}^{(3)1233} < 1.11,
  \end{aligned}
\end{equation}
with $tq+\overline{t}q$ production at the same energy leading to
somewhat weaker bounds.  The operator
$O^{(3)3123}_{qq}=-O^{(1)3123}_{qq}+$ terms with 0 or 2 top fields, so
the bounds on $C^{(3)3123}_{qq}$ from single-top production are the
same as those for $-C^{(1)3123}_{qq}$ in (\ref{eq:bnds.4a}).  Top pair
production is less sensitive than single-top processes to these
couplings, leading to bounds about twice as large as those in
(\ref{eq:bnds.4a}) at 8 TeV and larger at lower energies.  For the
operator $O^{(1)1233}_{qq}$, which does not contribute to single-top
production, the bounds obtained from $t\overline{t}$ production at 8
TeV are,
\begin{equation}
  \label{eq:bnds.4b}
    -4.72 < C_{qq\,r}^{(1)1233} < 4.58,
\quad
    -5.18 < C_{qq\,i}^{(1)1233} < 5.18,
\end{equation}
significantly weaker than the analogous limits in (\ref{eq:bnds.4a}).

The operators $O_{qq}^{(1)3313}$ and
$O_{qq}^{(3)3313}$ also contribute to both $tq$ and $tb$ production.
As shown in Figure \ref{fig:feyn5} (c), $tq$ production through these
operators involves two $b$ quarks in the initial state, leading to
very low sensitivity to these couplings.  More restrictive bounds are
furnished by $tb$ production (Figure \ref{fig:feyn6} (b)).  From the
cross section measurement at 2 TeV \cite{cdf14a} we get,
\begin{equation}
  \label{eq:bnds.5}
  |C_{qq}^{(1)3313}| < 4.92,
  \qquad
  |C_{qq}^{(3)3313}| < 2.57,
\end{equation}
Neither production channel, $tq$ or $tb$, possesses significant
sensitivity to $O_{qq}^{(1,3)3323}$.  

\begin{figure}[ht!]
  \centering
% fgr.4qa3.pd.pdf}  
\includegraphics[scale=0.88]{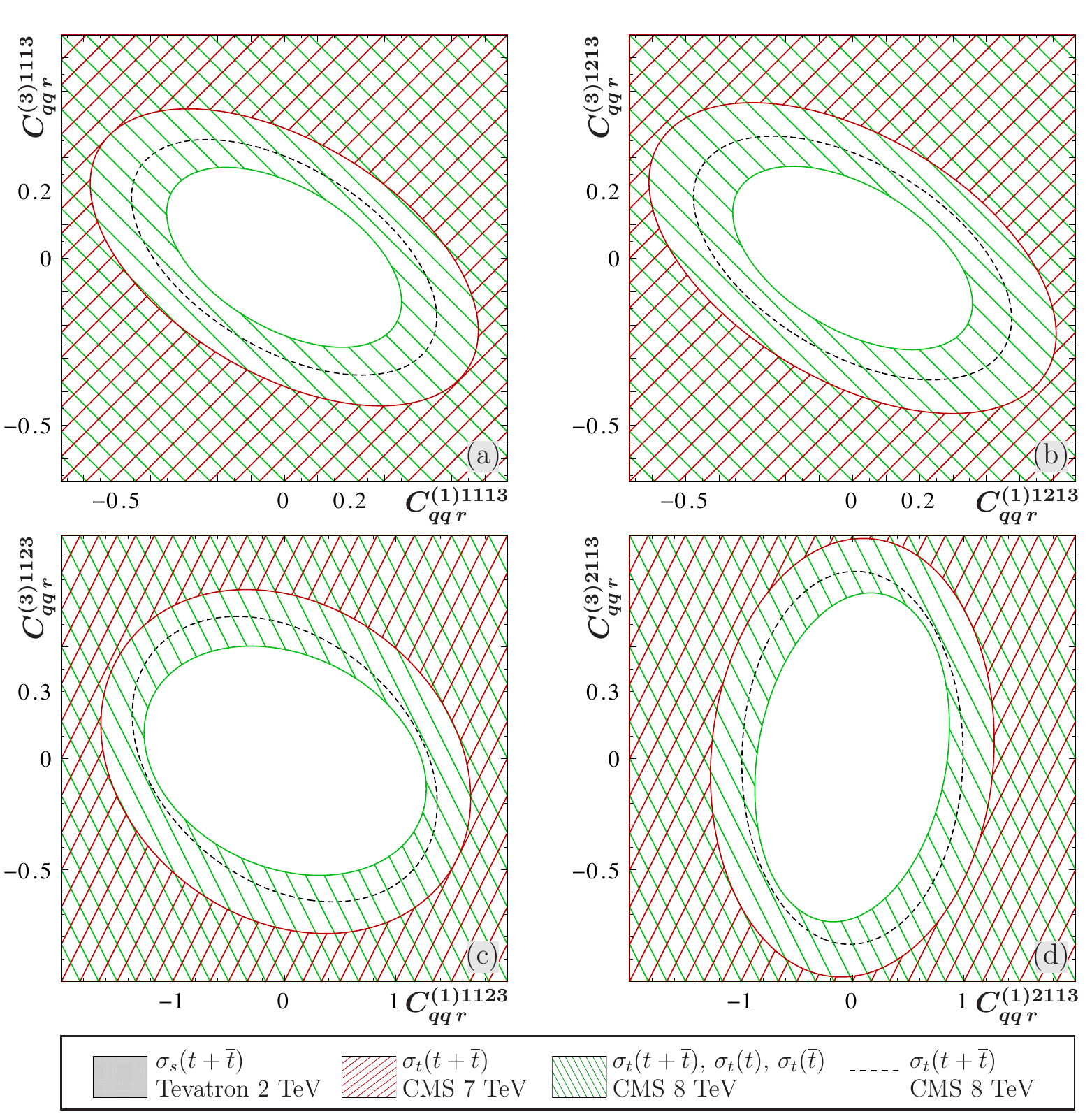}  
  \caption{Allowed parameter regions at 95\% CL for contact four-quark
effective couplings.  
Red hatched area: region excluded by the cross section
for $pp\rightarrow tq+\overline{t}q$ at 7 TeV \cite{cms2012}.
Green hatched area: region excluded by the cross sections for
$pp\rightarrow tq+\overline{t}q$, $tq$, $\overline{t}q$ at 8 TeV
\cite{cms2014}. Black dashed line: region excluded by the cross
section for $pp\rightarrow tq+\overline{t}q$ at 8 TeV alone
\cite{cms2014}. Gray area: region excluded by the
cross section for $p\overline{p}\rightarrow
t\overline{b}+\overline{t}b$ at 1.96 TeV \cite{cdf14a}. 
}
  \label{fig:fourfermion1}
\end{figure}

In Figure \ref{fig:fourfermion1} we show allowed regions for four
pairs of couplings $C^{(1)ijk3}_{qq\,r}/C^{(3)ijk3}_{qq\,r}$ with
$i,j,k<3$.  The most restrictive limits for these couplings are set in
all cases by the $tq$ production cross section at 8 TeV, with the
combined cross section for $tq+\overline{t}q$ yielding slightly weaker
bounds.  As also seen in the figure, there is sizeable interference
between the singlet-singlet and triplet-triplet amplitudes,
proportional to $C^{(1)}_{qq}$ and $C^{(3)}_{qq}$ respectively.

\begin{figure}[ht!]
  \centering
% fgr.ttx.pd.1.pdf
\includegraphics[scale=0.88]{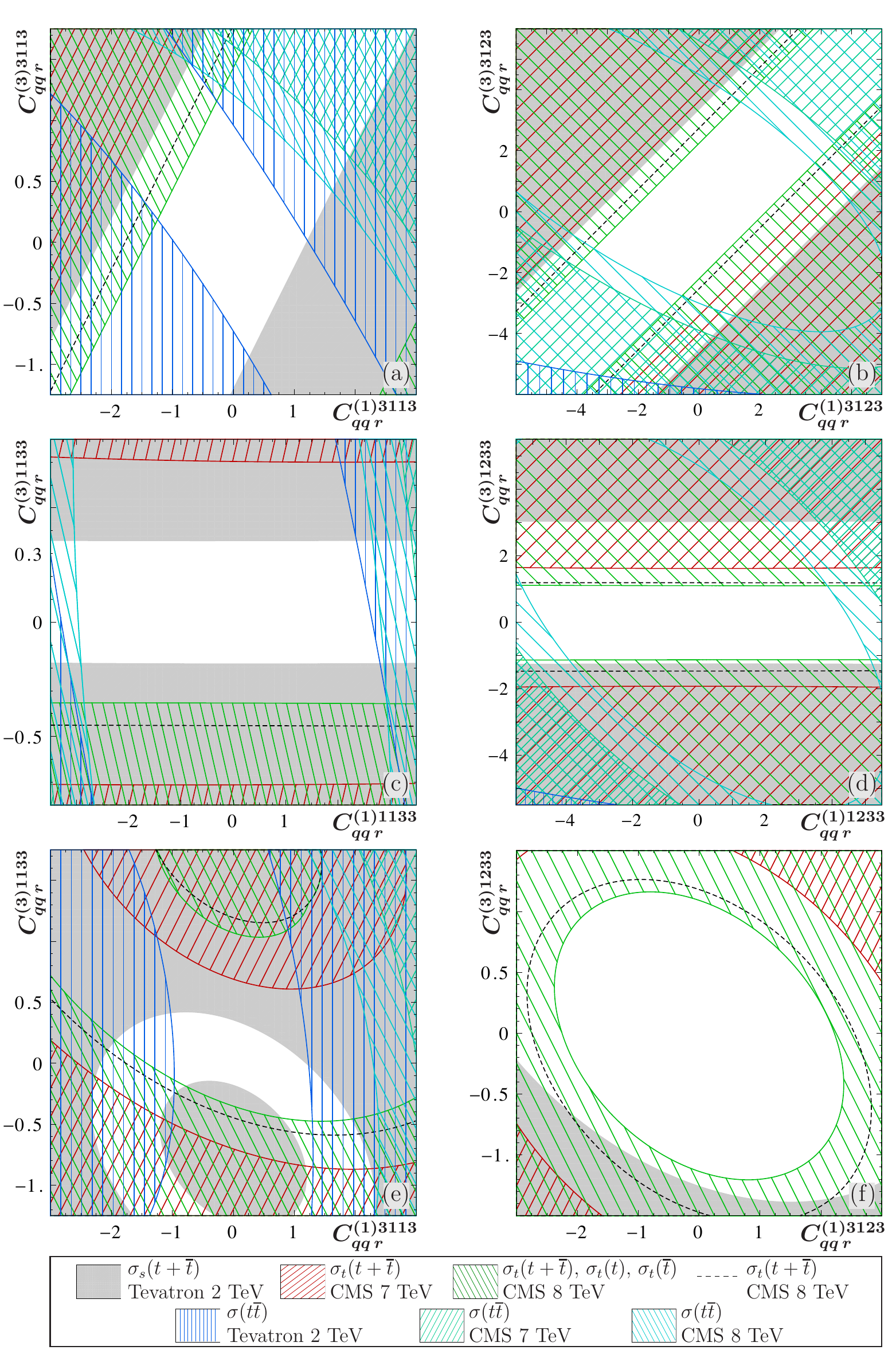}  
  \caption{Allowed parameter regions at 95\% CL for contact four-quark
effective couplings.  
Red and green hatched areas, gray area and black dashed line as in the
previous figure.  Regions excluded by the cross section for
$pp\rightarrow t\overline{t}$: blue hatched area (1.96 TeV
\cite{cdfd0ttx14}), light-green hatched area (7 TeV \cite{cmsttx12b}),
light-blue hatched area (8 TeV \cite{cmsttx13}). 
}
  \label{fig:fourfermion1.5}
\end{figure}

Figure \ref{fig:fourfermion1.5} displays allowed regions for six pairs
of four-quark couplings involving two third-generation quarks.  The
single-top cross sections do not depend on
$C^{(1)31k3}_{qq\,r}+C^{(3)31k3}_{qq\,r}$ or on $C^{(1)1k33}_{qq\,r}$,
$k=1,2$, as seen in Figure \ref{fig:fourfermion1.5} (a)--(d).  The
limits in those directions are set by the $t\overline{t}$ production
cross section.  We remark the fundamental role played by Tevatron data
in bounding the couplings involving only first- and third-generation
quarks (left column in the figure), whereas those involving one first-
and one second-generation quarks (right column in the figure) are
bounded by LHC 8 TeV data.

\begin{figure}[ht!]
  \centering
% fgr.4qb.pd.pdf
\includegraphics[scale=0.88]{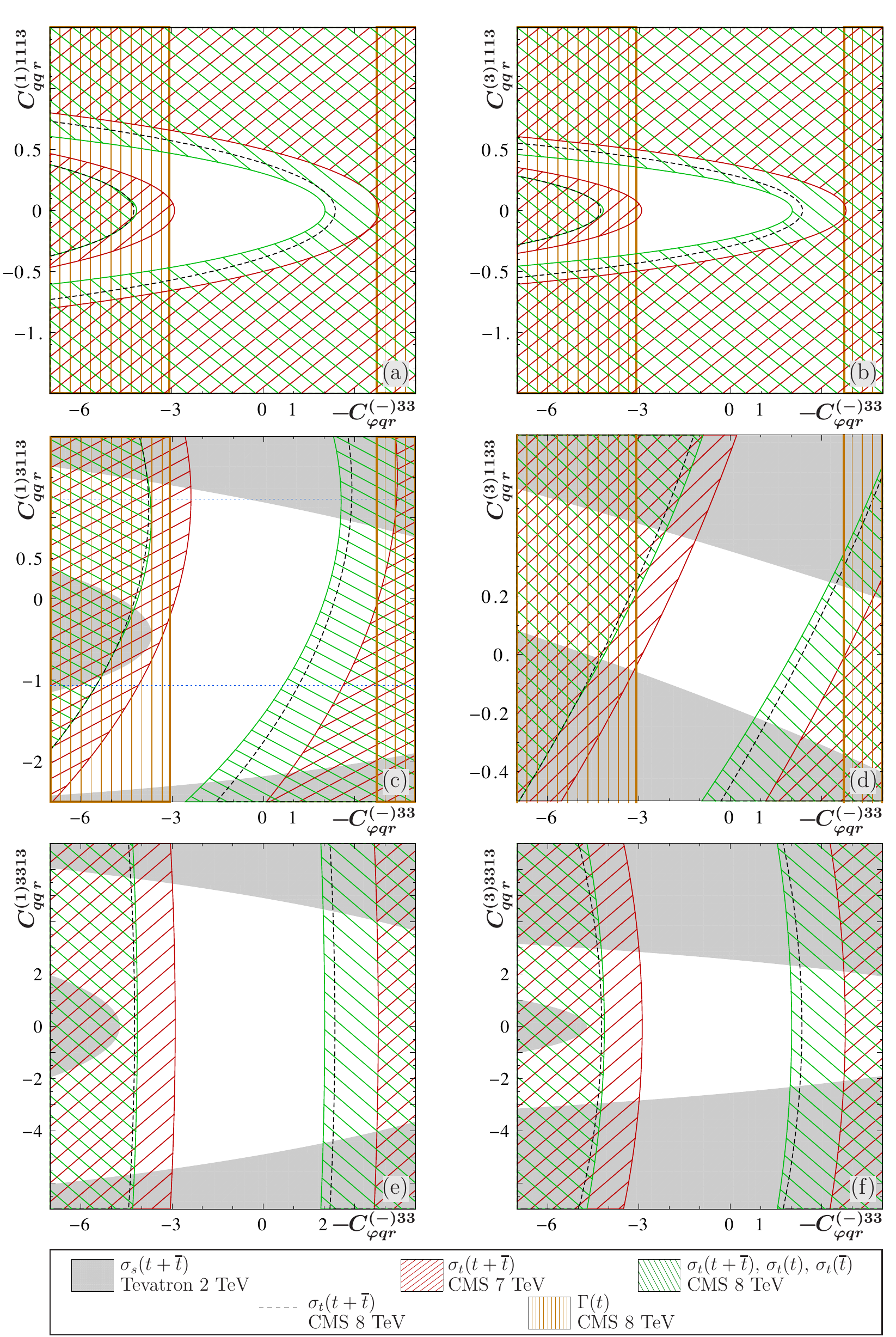}  
  \caption{Allowed parameter regions at 95\% CL for four-quark
effective couplings versus the flavor-diagonal left-handed vector
effective coupling. 
Red and green hatched areas, gray area and black dashed line as in 
figure \ref{fig:fourfermion1}.  Orange hatched area: region excluded by
the top decay width \cite{cmstwqratio}. Blue dotted lines: region
excluded by the $t\overline{t}$ production cross section at 1.96 TeV
\cite{cdfd0ttx14}. 
}
  \label{fig:fourfermion2}
\end{figure}

In Figure \ref{fig:fourfermion2} we show the allowed regions in the
plane of the four-quark couplings $C_{qq}$ involving first-generation
quarks and the flavor-diagonal vector coupling $-C_{\varphi
  qr}^{(-)33}$ (i.e., the parameter $V_L$).  The importance of $tb$
production to bound those couplings involving more than one
third-generation quark is apparent from the four lower panels
though,  as seen in the figure, in the case of $C^{(1)3113}_{qq\,r}$
more restrictive bounds result from $t\overline{t}$ production.

\begin{figure}[ht!]
  \centering
% fgr.4qcm.pd.pdf  
\includegraphics[scale=0.85]{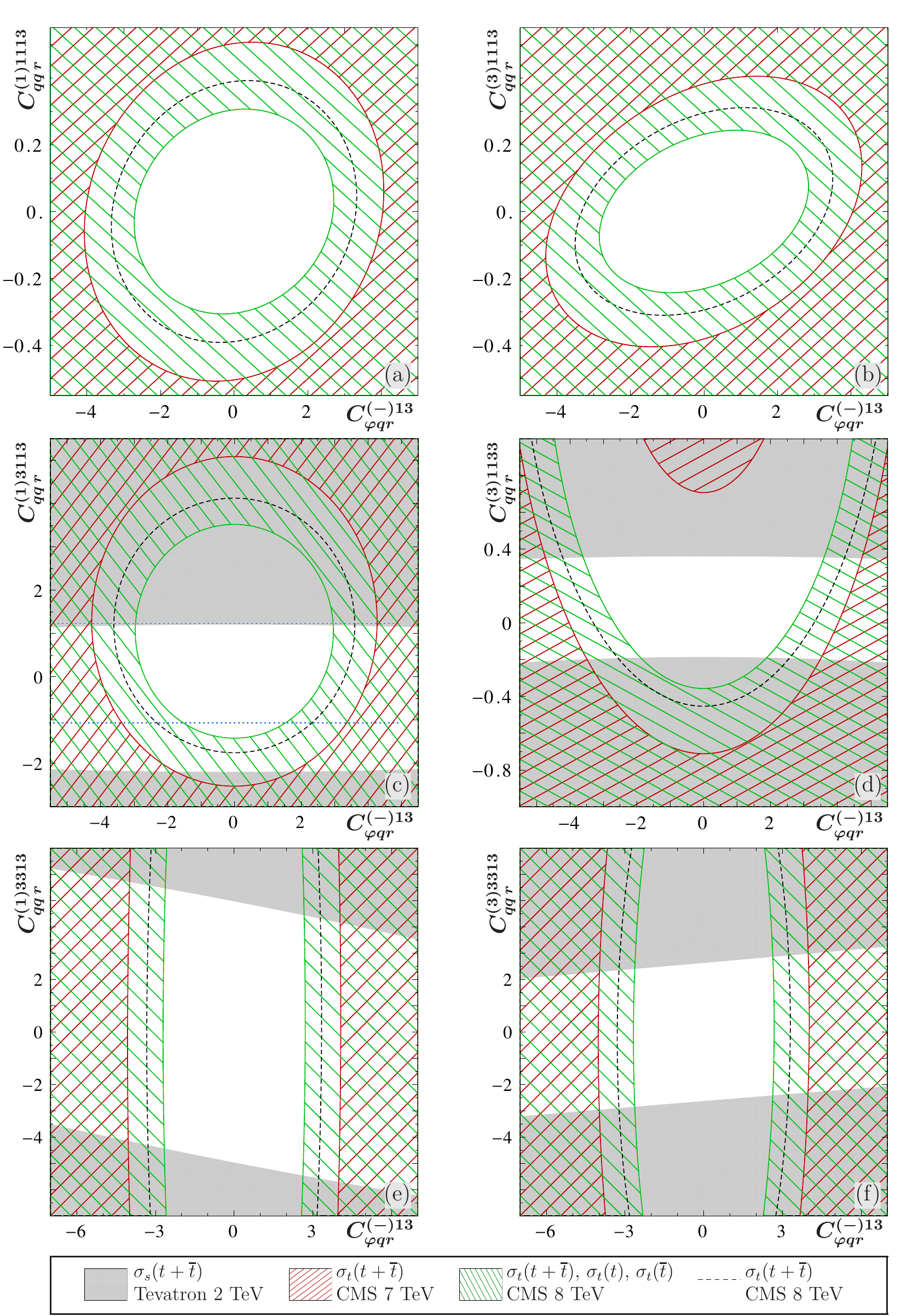}  
  \caption{Allowed parameter regions at 95\% CL for four-quark
effective couplings versus the flavor off-diagonal left-handed vector 
effective coupling. Color codes as in  figure
\ref{fig:fourfermion1}. Blue dotted lines: region excluded by the
$t\overline{t}$ production cross section at 1.96 TeV
\cite{cdfd0ttx14}.   
}
  \label{fig:fourfermion3}
\end{figure}

In Figure \ref{fig:fourfermion3} we show the allowed region on the
plane of the same four-quark coupling as in the previous figure and
the flavor off-diagonal vector coupling $C_{\varphi qr}^{(-)13}$.  The
interference between the amplitudes proportional to
$C^{(1,3)1113}_{qq}$ and those proportional to $C_{\varphi
  qr}^{(-)13}$, as well as between the amplitudes proportional to
$C^{(1)3113}_{qq}$ , $C^{(3)1133}_{qq}$ and the SM ones is clearly
seen in the figure.

%%%%%%%%%%%%%%%%%%%%%%%%%%%%%%%%%%%%%%%%%%%%%%%%%%%%%%%
%%%%%%%%%%%%%%%%%%%%%%%%%%%%%%%%%%%%%%%%%%%%%%%%%%%%%%%%%%%%%%%%%%

\section{Conclusions}
\label{conclusions}

In this paper we have obtained limits on $Wtq$ vertices in the context
of the $SU(2)\times U(1)$-gauge invariant effective Lagrangian of
dimension six.  We worked with the basis of operators listed in
\cite{aguilaroperators,rosiek}, with the operator normalization used
in \cite{maltoni15,zha14}.  In the SM the $Wtq$ couplings are
suppresed by the CKM parameters.  No precise direct measurements of
$V_{td}$, $V_{ts}$ exist so far, but there are studies that propose to
use single top production distributions in order to achieve higher
accuracy \cite{aguilarckm}.  In this study we refer to the $Wtq$
vertices as generated by the dimension six operators.  There are
previous studies on limits for the diagonal anomalous $Wtb$ coupling
based on single top production and $W$-helicity fractions in the $t\to
bW$ decay, with \cite{fabbrichesi,onofre,whelicity2,buckley} the most
recent references.  However, no similar direct limits have been
reported before for the flavor off-diagonal $Wts$ and $Wtd$ couplings.
There are 4 independent dimension six operators that give rise to
$Wtq$ vertices: $O_{\varphi q}^{(3)k3}$, $O_{\varphi ud}^{3k}$,
$O_{uW}^{k3}$ and $O_{dW}^{3k}$.  Three of them generate simultaneusly
neutral current couplings.  Only $O_{\varphi ud}^{3k}$ generates a CC
coupling exclusively, which is the right-handed vector $W_\mu^-
\overline{d}_R \gamma_\mu t_R$.  For the other three operators, we
have followed the strategy used in Ref.~\cite{maltoni15} and we have
defined six linear combinations with other three operators $O_{\varphi
  q}^{(1)k3}$, $O_{uB}^{k3}$ and $O_{dB}^{3k}$, so as to define
separately $Ztq_u$, $Atq_u$, $Zbq_d$ and $Abq_d$ interactions, with
$q_u$, $q_d$ any up- or down-type quark.

In order to obtain bounds on these six operators we have considered
the FCNC processes $b\to d\gamma,s\gamma$ and $Z\to bd,bs$, the CC
decays $t\to Wq$ (through its total width, branching fractions and
$W$-helicity fractions) and the single-top production processes $pp\to
tq$ and $p\overline{p}\to tb$.  These results are summarized in Table
\ref{fcnclimits} and in Figures
\ref{fig:offdiagonal2}--\ref{fig:diag-offdiag}.  For the operators
$O^{(+)k3}_{\varphi q}$, $O_{dA}^{3k}$ and $O_{dZ}^{3k}$ ($k=1,2$),
involving bottom-strange and bottom-down quark interactions, we find
that the best bounds are obtained from the LEP measurement of $Z\to
bq$ and the most recent experimental result on the $B\to X_q \gamma$
decay ($q=d,s$).  The direct bounds on these operators are obtained
here for the first time.  Notice, however, that for
$O^{(+)k3}_{\varphi q}$ there are stronger indirect bounds
\cite{soreq}.  We obtain bounds for the operators $O_{\varphi
  ud}^{3k}$ ($k=1,2$), also for the first time.  The best bounds on
$O_{\varphi ud}^{31}$ result from the single-top production cross
section at 8 TeV, and on $O_{\varphi ud}^{32}$ from the ratio of top
branching fractions $Br(t\to tb)/\sum Br(t\to tq)$.  We also show in
the table and figures, for completeness, the best bounds reported in
\cite{maltoni15} on $O^{(-)k3}_{\varphi q}$, $O^{k3}_{uZ}$, from $t\to
jZ$, and on $O^{k3}_{uA}$ from $gq\to t\gamma$. For the
flavor-diagonal effective $Wtb$ coupling we have made an improvement
of the previous analyses \cite{fabbrichesi,onofre,whelicity2,buckley}
using the most recent experimental results on $W$-helicity fractions
in top quark decay from $t\overline{t}$ production at the LHC
\cite{cmswhelicity15}.

We have considered also contact-interaction operators involving the
top quark, focusing on those four-quark operators related to the $Wtq$
ones by the SM equations of motion.  Our results are given in section
\ref{sec:contact} and in Figures
\ref{fig:fourfermion1}--\ref{fig:fourfermion3}.  The flavor off-diagonal
operators $O_{qq}^{(1,3)ijk3}$ (with $ijk=111$ or a permutation of
$112$) involving three light quarks and the top are considered here
for the first time.  The single-top production process $pp\to tq$
measured at the LHC possesses strong sensitivity to these operators,
resulting in the tight bounds on the associated couplings reported
above.  In fact, the bounds on $C_{qq}^{(1,3)1113}$,
$C_{qq}^{(1,3)1213}$ (equation (\ref{eq:bnds.1113}) and Figures
\ref{fig:fourfermion1}--\ref{fig:fourfermion3}) are the strongest ones
found in this paper for interactions vertices involving the top quark.
The flavor off-diagonal operators $O_{qq}^{(1,3)3313}$ had not been
considered before in the literature.  For this coupling it is the
single-top process $p\overline{p}\to tb$ measured at the Tevatron that
has some sensitivity, leading to the bounds in equation
(\ref{eq:bnds.5}).  The flavor-diagonal triplet operator
$O_{qq}^{(3)1133}$ had already been discussed in \cite{fabbrichesi},
though not the singlet $O_{qq}^{(1)3113}$.  Both operators lead to
interference with the SM, stronger for the triplet operator.  The
sensitivity to these couplings comes mostly from the Tevatron result
for $p\overline{p}\to tb$ production.  Our bounds on
$C_{qq}^{(3)1133}$ (equation (\ref{eq:bnds.4}) and Figures
\ref{fig:fourfermion1}--\ref{fig:fourfermion3}) are somewhat tighter
than those reported in \cite{fabbrichesi} for the reasons explained at
the end of section \ref{sec:statanal}.

Single top production at the LHC will mostly have a direct impact on
the limits for four-fermion quark operators as well as the
flavor-diagonal couplings $C^{(\pm)33}_{\varphi q}$ and flavor
off-diagonal $C^{3k}_{\varphi ud}$ of top-gauge boson couplings.
Also, $W$-helicity fractions will set strong constraints on the other
diagonal $Wtb$ couplings.  On the other hand, FCNC processes like
$t\to jZ$ and $pp\to t\gamma$ \cite{maltoni15} at the LHC will be the
best options to set strong constraints to the operators that give rise
to the off-diagonal $C^{(-)k3}_{\varphi q}$, $C^{k3}_{uZ/A}$ and
$C^{3k}_{dZ/A}$ couplings. 

\paragraph*{Acknowledgements}

We acknowlegde support from Conacyt and Sistema Nacional de
Investigadores de M\'exico.  We thank Gauthier Durieux and Cen Zhang
for useful comments.

\end{document}